\pdfoutput=1

\documentclass[10pt,aps,pre, twocolumn,superscriptaddress,notitlepage, longbibliography]{revtex4-2}

\usepackage{amsmath}
\usepackage{amssymb}
\usepackage{graphicx}
\usepackage{bm}

\usepackage{txfonts}

\hbadness=\maxdimen
\vbadness=\maxdimen
\def\be{\begin{equation}}
\def\ee{\end{equation}}
\def\bea{\begin{eqnarray}}
\def\eea{\end{eqnarray}}
\def\ba{\begin{array}}
\def\ea{\end{array}}

\usepackage{xcolor}
\usepackage[colorlinks=true,allcolors=blue]{hyperref}

\allowdisplaybreaks

\begin{document}

\title{Nonlinear Poisson effect in affine semiflexible polymer networks}

\author{Jordan L.\ Shivers}
\altaffiliation{Present affiliation: James Franck Institute and Department of Chemistry, University of Chicago, Chicago, Illinois 60637, USA}
\affiliation{Department of Chemical and Biomolecular Engineering, Rice University, Houston, TX 77005, USA}
\affiliation{Center for Theoretical Biological Physics, Rice University, Houston, TX 77005, USA}
\affiliation{James Franck Institute, University of Chicago, Chicago, IL 60637, USA}
\affiliation{Department of Chemistry, University of Chicago, Chicago, IL 60637, USA}
\author{Fred C.\ MacKintosh}
\affiliation{Department of Chemical and Biomolecular Engineering, Rice University, Houston, TX 77005, USA}
\affiliation{Center for Theoretical Biological Physics, Rice University, Houston, TX 77005, USA}
\affiliation{Department of Chemistry, Rice University, Houston, TX 77005, USA} 
\affiliation{Department of Physics \& Astronomy, Rice University, Houston, TX 77005, USA}

\begin{abstract}
    Stretching an elastic material along one axis typically induces contraction along the transverse axes, a phenomenon known as the Poisson effect. From these strains, one can compute the specific volume, which generally either increases or, in the incompressible limit, remains constant as the material is stretched. However, in networks of semiflexible or stiff polymers, which are typically highly compressible yet stiffen significantly when stretched, one instead sees a significant reduction in specific volume under finite strains. This volume reduction is accompanied by increasing alignment of filaments along the strain axis and a nonlinear elastic response, with stiffening of the apparent Young's modulus. For semiflexible networks, in which entropic bending elasticity governs the linear elastic regime, the nonlinear Poisson effect is caused by the nonlinear force-extension relationship of the constituent filaments, which produces a highly asymmetric response of the constituent polymers to stretching and compression. The details of this relationship depend on the geometric and elastic properties of the underlying filaments, which can vary greatly in experimental systems. Here, we provide a comprehensive characterization of the nonlinear Poisson effect in an affine network model and explore the influence of filament properties on essential features of the macroscopic response, including strain-driven alignment and volume reduction.
\end{abstract}


\maketitle

Networks of semiflexible biopolymers provide elasticity to many biological materials, which they imbue with a variety of distinctive properties that set them apart from conventional elastic media \cite{kasza_cell_2007,pritchard_mechanics_2014,broedersz_modeling_2014,burla_mechanical_2019}. One of the more remarkable features of biopolymer networks is their highly \textit{asymmetric} mechanical response: they tend to respond to increasing deformation with a strongly increasing stiffness, often by more than an order of magnitude \cite{fung_elasticity_1967, gardel_elastic_2004, storm_nonlinear_2005}, while providing a comparatively weak or even softening response to compression \cite{van_oosten_uncoupling_2016,xu_compressive_2017,vahabi_elasticity_2016,van_oosten_emergence_2019,ed-daoui_poroviscoelasticity_2021}. The stiffness of typical elastic materials, in contrast, usually depends weakly on the magnitude or nature of the applied load. A striking consequence of the asymmetric response of biopolymer networks is seen in the manner in which they exhibit the \textit{Poisson effect}, which refers to the tendency of a material stretched along one axis to contract along the transverse axes. Unlike ordinary materials, for which this effect depends only mildly on strain, biopolymer networks exhibit a strongly nonlinear Poisson effect under finite extensional strain \cite{brown_multiscale_2009,vader_strain-induced_2009,steinwachs_three-dimensional_2016,ban_strong_2019}.

\begin{figure}[b]
    \includegraphics[width=1\columnwidth]{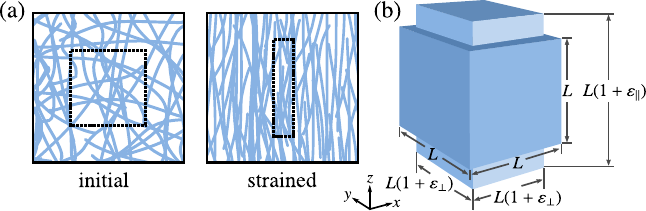}
    \caption{\label{fig1} Extensile strain drives alignment and densification of networks of crosslinked semiflexible polymers. (a) A schematic of the nonlinear Poisson effect in biopolymer networks. Applying (vertical) extensile strain $\varepsilon_\parallel$ to an initially isotropic semiflexible polymer network (left) and allowing the (horizontal) transverse strain $\varepsilon_\perp$ to freely vary results in alignment of polymers along the vertical axis, along with local densification due to the transverse contraction of the network (right). The dashed box represents the same portion of the network in the initial and strained states. In the strained state, the stress along the (vertical) strain axis, $\sigma_{\parallel}$ is positive, whereas the transverse (horizontal) stress is relaxed, $\sigma_{\perp} = 0$. (b) Sketch of a 3D cubic volume element of initial length $L$ before and after the application of extensile strain $\varepsilon_{\parallel}>0$ along the $z$-axis. If the Poisson's ratio $\nu$ is positive, requiring $\sigma_{\perp} = 0$ results in transverse contraction quantified by strains $\varepsilon_{\perp} < 0$.}
\end{figure}

In the small strain limit, this effect is quantified by Poisson's ratio  $\nu = \lim_{\varepsilon_\parallel\to 0^+}[-\varepsilon_\parallel/\varepsilon_\perp]$, in which $\varepsilon_\parallel$ is the applied extensional strain and $\varepsilon_\perp$ is the resulting transverse strain in the absence of transverse stress \cite{poisson_note_1827, love_treatise_1906}. As with any stable, isotropic, three-dimensional material, the true (small strain) Poisson's ratio of biopolymer networks is strictly constrained within the range $\nu \in [-1, 1/2]$ \cite{greaves_poissons_2011}. However, under finite applied (extensional) strains, biopolymer networks in various contexts have been shown to exhibit an unusually large and highly strain-dependent \textit{apparent} Poisson's ratio \cite{brown_multiscale_2009,vader_strain-induced_2009,steinwachs_three-dimensional_2016,ban_strong_2019}, corresponding to significant contraction along the transverse axes in response to small increases in extension. This behavior, sketched in Fig. \ref{fig1}, produces a contracted state with pronounced filament alignment along the strain axis \cite{vader_strain-induced_2009} and a significant decrease in the local volume occupied by the network \cite{brown_multiscale_2009,steinwachs_three-dimensional_2016} that is accompanied by an expulsion of solvent and a significant reduction in the average pore size \cite{vader_strain-induced_2009}. In contrast, the apparent Poisson's ratio changes very little under compression. Although the nonlinear Poisson effect observed in biopolymer networks is a fundamentally network-scale behavior, it nonetheless reflects the mechanical asymmetry of the underlying components: the individual filaments resist finite tension significantly more strongly than they resist finite compression.

The large configurational changes caused by the nonlinear Poisson effect could have meaningful biological consequences, e.g. in cell migration, mechanosensing, and intracellular transport. For example, the Poisson effect produces significant changes in network porosity, which in the extracellular matrix can strongly influence the motility of migrating cells \cite{wolf_physical_2013, paul_cancer_2017}. Likewise, the Poisson effect leads to the strain-driven formation of dense, highly oriented network regions between contractile cells \cite{vader_strain-induced_2009,ban_strong_2019} which can influence mechanosensing \cite{abhilash_remodeling_2014,wang_long-range_2014,hall_fibrous_2016} and induce directed cell motion \cite{notbohm_microbuckling_2015, sengupta_principles_2021}. Within cells, changes in cytoskeletal network orientation and density caused by the Poisson effect could likewise influence the transport of intracellular cargo both by passive diffusion and by active (e.g., molecular motor-driven or polymerization-driven) transport processes \cite{ahmed_active_2014}. The influence of the Poisson effect on filament orientation could also play an important role in other active mechanical processes involving the cytoskeleton, such as actomyosin-based cell motility \cite{svitkina_analysis_1997,keren_biophysical_2008} and the organization of stress fibers in response to applied load \cite{lee_cyclic_2010,nagayama_strain_2012,kaunas_dynamic_2015,kumar_filamin_2019}. Developing a more comprehensive understanding of how the nonlinear Poisson effect is controlled, e.g. by filament stiffness and network structure, is therefore of interest both for understanding existing biological systems and for designing reconstituted or synthetic materials that mimic these behaviors.

To date, studies addressing the Poisson effect in biopolymer gels have focused primarily on models relevant to networks of stiff athermal biopolymers, such as thick fibrin or collagen fibers, that exhibit enthalpic elasticity and tend to deform in a highly non-affine (inhomogeneous) manner \cite{kabla_nonlinear_2007,steinwachs_three-dimensional_2016,picu_poissons_2018,ban_strong_2019,shivers_nonlinear_2020}. While such models are appropriate for the stiff constituents of the extracellular matrix, many biopolymers, particularly on the subcellular scale, are semiflexible and thus exhibit entropic elasticity under small strains \cite{mackintosh_elasticity_1995,odijk_stiff_1995}. For networks of semiflexible polymers, mechanical models neglecting non-affinity have been an effective starting point to capture the mechanics of reconstituted gels, even in highly nonlinear regimes \cite{gardel_elastic_2004,storm_nonlinear_2005}, but our understanding of the nonlinear Poisson effect in these systems remains limited. In this work, we aim to elucidate the mechanism underlying the Poisson effect in such systems. Specifically, we consider an affine model of a network of semiflexible filaments, in which the coordinates of the crosslinks (or entanglements) transform affinely with the macroscopic applied strain \cite{treloar_photoelastic_1954,treloar_physics_1975,treloar_non-gaussian_1979,wu_improved_1993,wu_network_1995,boyce_constitutive_2000,palmer_constitutive_2008,storm_nonlinear_2005,cioroianu_normal_2013,meng_theory_2017,song_hyperelastic_2022,rubinstein_polymer_2006}. We systematically characterize the strain-dependent mechanical and configurational behavior of initially isotropic, three-dimensional networks under varying applied uniaxial strain with a zero transverse stress condition. These conditions mimic a typical experiment used to measure a material's Poisson's ratio. Using this model, we explore the influence of various filament properties, including bending rigidity, stretch modulus, and contour length on the nonlinear Poisson effect and the associated strain dependence of the incremental Poisson's ratio, nematic alignment, relative volume, and stress.

The manuscript is organized as follows. In Sec. \ref{sec:forceextension}, we briefly describe the force-extension relationship of individual filaments, for which we adopt an existing model of stretchable wormlike chains. Sec. \ref{sec:affinemodel} describes the affine network model, into which the force-extension relationship is fed as an input, and define the associated stress and nematic tensors. In Sec. \ref{sec:strainprotocol}, we describe the modeled uniaxial strain protocol and the calculation of the strain-dependent stress and filament alignment. In Sec. \ref{sec:results}, we present and discuss our results.

\section{Model}

\subsection{Force-extension relations} \label{sec:forceextension}

We consider a simple mechanical model of extensible wormlike chains (WLCs) \cite{odijk_stiff_1995, storm_nonlinear_2005}, with which we can parametrically tune the filament force-extension relationship between the purely enthalpic Hookean spring limit (i.e., a linear force-extension relationship with spring constant $\mu/\ell_c$) and the highly nonlinear limit of inextensible wormlike chains with entropic linear elasticity governed by a bending rigidity $\kappa$ \cite{mackintosh_elasticity_1995, gardel_elastic_2004}. Between these regimes, the model captures the more generic behavior of extensible wormlike chains, filaments that exhibit entropic linear elasticity governed by a bending rigidity $\kappa$ but, under sufficient extension, cross over to an enthalpic stretching regime governed by the stretch modulus $\mu$.  

To obtain the force-extension relationship for extensible wormlike chains, we first need to derive the corresponding relationship for the inextensible limit. Consider an inextensible wormlike chain filament of contour length $\ell_c$ and persistence length $\ell_p$ under applied tension $\tau$ at thermal equilibrium. Thermal fluctuations excite the filament's bending modes, reducing the ensemble-averaged end-to-end length $\ell_\mathrm{WLC}(\tau)$ with respect to the contour length \cite{vahabi_normal_2018,mackintosh_elasticity_1995}. Specifically, 
\begin{align}
\begin{split}
\ell_\mathrm{WLC}(\tau)  & = \ell_c - \frac{\ell_c^2}{\pi^2\ell_p}\sum^\infty_{n=1}{\frac{1}{n^2+\tau\ell_c^2/(\pi^2\kappa)}} \\
& = \ell_c - \frac{\kappa}{2\tau \ell_p}\left(\frac{\ell_c \sqrt{\tau/\kappa}} {\tanh{\left(\ell_c \sqrt{\tau/\kappa}\right)}}-1\right)
\end{split}
\end{align}
in which $\kappa = k_BT\ell_p$ is the filament's bending rigidity and the sum is taken over all mode numbers.  Under zero applied tension ($\tau = 0$), the ensemble-averaged end-to-end length is $\ell_0 \equiv \ell_\mathrm{WLC}(\tau = 0)  = \ell_c - \ell_c^2/(6\ell_p)$. We can thus write the extension-force relation $\delta\ell_\mathrm{WLC}(\tau) = \ell_\mathrm{WLC} - \ell_0$ for an inextensible wormlike chain as
\begin{equation}
\delta \ell_\mathrm{WLC}(\tau) =  \frac{\ell_c^2}{6\ell_p} \left( 1 - \frac{3\kappa}{\ell_c^2\tau }\left(\frac{\ell_c \sqrt{\tau/\kappa}} {\tanh{\left(\ell_c \sqrt{\tau/\kappa}\right)}}-1\right)\right)
\end{equation}
The above relation assumes the filaments are inextensible, such that $\tau$ diverges as the thermally contracted filament is stretched to its contour length, $\delta\ell_\mathrm{WLC} \to \ell_c^2/(6\ell_p)$. Real filaments are of course not entirely inextensible, and with sufficient strain they instead transition from the entropic regime to an enthalpic  regime governed by a stretching modulus $\mu$ \footnote{For a homogeneous cylindrical elastic rod of radius $r$, the stretching modulus $\mu$ is related to the bending stiffness $\kappa$ as $\mu = 4\kappa/r^2$ \cite{storm_nonlinear_2005}.}. To capture this behavior, we define a more general extension-force relation $\delta\ell(\tau)$ for extensible wormlike chains by renormalizing $\delta\ell_\mathrm{WLC}$ as follows \cite{odijk_stiff_1995,storm_nonlinear_2005,broedersz_modeling_2014} :
\begin{equation}\label{eq:force_extension}
\delta\ell(\tau)=\ell_c \tau/\mu + \delta\ell_\mathrm{WLC}\left(\tau\left[1+\tau/\mu\right]\right)
\end{equation}
Note that taking $\mu\to\infty$ with finite $\kappa$ recovers the inextensible wormlike chain limit, $\delta\ell_\mathrm{WLC} (\tau) = \lim_{\mu \to \infty} \delta\ell(\tau)$, while taking $\kappa\to\infty$ with finite $\mu$ produces a simple linear Hookean relationship. 

The force-extension relationship is essentially controlled by two dimensionless numbers: the dimensionless bending rigidity, $\tilde{\kappa} \equiv \kappa/(\mu \ell_c^2)$, which compares the relative strengths of the bending and stretching resistance of the filaments, and a dimensionless persistence length $\tilde{\ell}_p \equiv \ell_p/\ell_c$, which compares the filament persistence length $\ell_p$ to the contour length $\ell_c$, the latter of which describes the backbone length of the filament between constraints (crosslinks or entanglements). The dimensionless bending rigidity $\tilde{\kappa}$ controls the degree to which the force-extension relation is nonlinear; the Hookean limit corresponds to $\tilde{\kappa} \to \infty$, whereas the the inextensible limit corresponds to $\tilde{\kappa}\to 0$. The dimensionless persistence length $\tilde{\ell}_p$ controls the degree to which thermal fluctuations reduce the rest length $\ell_0$ of the filament relative to the full contour length $\ell_c$, as $\ell_0 = \ell_c ( 1 - 1 / (6\tilde{\ell}_p))$. Equivalently, $\tilde{\ell}_p$ governs the level of extensional strain (applied along the filament end-to-end vector) required to bring the initially contracted filament to its contour length,
\begin{equation} \label{eq:critical_extension}
\varepsilon_c = \frac{\ell_c - \ell_0}{\ell_0} = \frac{1}{6 \tilde{\ell}_p - 1}.
\end{equation} 
We will refer to $\varepsilon_c$ as the ``critical extension,'' at which the resulting tension $\tau$ diverges in the inextensible ($\tilde{\kappa}\to 0$) limit or becomes stretching-dominated ($\propto \mu$) for filaments with finite $\tilde{\kappa} > 0$.

\begin{figure}[htb!]
    \includegraphics[width=1\columnwidth]{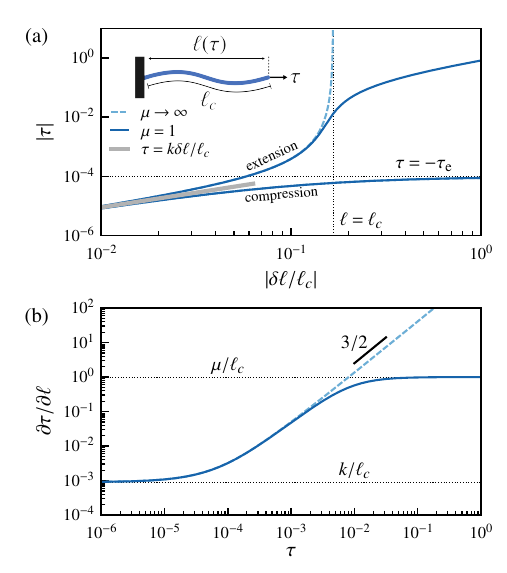}
    \caption{\label{fig:force_extension}
    A summary of the asymmetric force-extension relationship of semiflexible polymers: the wormlike chain model. (a) Force-extension relationships for inextensible ($\mu\to\infty$) and extensible ($\mu = 1$) wormlike chains with bending rigidity $\kappa = 10^{-5}$ and contour and persistence lengths $\ell_c=\ell_p$, under compression (dashed curves, $\delta\ell < 0$) and extension (solid curves, $\delta\ell > 0$), in which $\delta\ell = \ell - \ell_0$ is the difference between the end-to-end distance $\ell$ and the rest length $\ell_0$, defined as the thermally contracted average length under zero tension, $\ell_0 = \ell_c - \ell_c^2/(6\ell_p)$. As an inextensible ($\mu\to\infty$) wormlike chain is pulled to its full contour length $\ell \to \ell_c$, as indicated by the vertical dotted line, the tension $\tau$ diverges. The thick solid grey line indicates the linear relationship $\tau = k\delta\ell/\ell_c$ with $k = 1/(\mu^{-1} + \ell_c^3/(90\kappa\ell_p))$. The horizontal  dotted line indicates the Euler buckling force, $\tau_\mathrm{e} = -\kappa\pi^2/\ell_c^2$. (b) Chain stiffness $\partial \tau/\partial \ell$ as a function of applied tension $\tau$. For small tensions, $\partial \tau/\partial \ell \to k/\ell_c$. In the inextensible limit ($\mu\to\infty$), the stiffness scales as $\partial \tau/\partial \ell \propto \tau^{3/2}$ for large $\tau$. For finite $\mu$, one finds $\partial\tau/\partial\ell \to \mu/\ell_c$ for large $\tau$.}
\end{figure}

Having defined the extension-force relationship $\delta\ell(\tau)$, we can obtain the corresponding force-extension relationship $\tau(\delta\ell)$ by numerical inversion \cite{storm_nonlinear_2005}. In Fig. \ref{fig:force_extension}a, we plot the absolute value of the tension $|\tau|$ as a function of the normalized elongation  $|\delta\ell/\ell_c|$ for extensible ($\mu = 1$) and inextensible ($\mu \to \infty$) filaments under both extension and compression, with $\ell_p = \ell_c = 1$ ($\tilde{\ell}_p=1$) and $\kappa = 10^{-5}$. 
For small $\delta \ell/\ell_c$, the force-extension relationship exhibits a simple linear dependence $\tau = k \delta\ell/\ell_c$, in which the effective stiffness $k$ is given by $k = 1/(\mu^{-1} + \ell_c^3/(90\kappa\ell_p))$. Note that for the Hookean limit, this yields $k=\mu$, and for the inextensible WLC limit, $k=90\kappa\ell_p/\ell_c^3$ \cite{mackintosh_elasticity_1995,vahabi_normal_2018}. As $\kappa = 10^{-5}$ for the plotted curves, the values of $k$ for the inextensible and extensible cases are virtually identical. 
Under extension ($\delta\ell > 0$), the tension $\tau$ stiffens dramatically as the filament length $\ell$ approaches and, in the extensible case, exceeds the contour length $\ell_c$. For $\mu \to \infty$, $\tau$ diverges as $\ell \to \ell_c$, whereas for $\mu = 1$, the filament crosses over to a regime in which the tension is dominated by the stretch modulus, with  $\tau \propto \mu$. In Fig. \ref{fig:force_extension}b, we plot the filament stiffness $\partial\tau/\partial\ell$ as a function of the tension $\tau$. In the limit of small $\tau$, the stiffness for both extensible and inextensible filaments is given by $\partial \tau/\partial \ell = k/\ell_c$, whereas in the limit of large $\tau$, the stiffness of extensible filaments reaches an upper limit of $\partial\tau /\partial\ell = \mu/\ell_c$, while the stiffness of inextensible filaments exhibits a power law dependence on the tension with $\partial\tau/\partial\ell \propto \tau^{3/2}$ \cite{bustamante_entropic_1994,fixman_polymer_1973}.
Under compression ($\delta\ell < 0$), $\tau$ approaches a limiting value of $\tau = \tau_e$ (in the small $\tilde{\kappa}$ limit), in which $\tau_e = -\kappa \pi^2/\ell_c^2$ is the critical compressive force for Euler buckling of a rod of length $\ell_c$ and bending rigidity $\kappa$ \cite{broedersz_modeling_2014}.

\subsection{Affine network model} \label{sec:affinemodel}

We model the network as a collection of independent filament segments with random (isotropically distributed) initial orientations. The positions of the filament endpoints are assumed to transform affinely according to the macroscopically applied strain, and the resulting tension is assumed to act along the transformed end-to-end vector. The network-level stress is then obtained by averaging over all filament orientations in the deformed configuration. This model dates back to early work on rubber \cite{james_theory_1943,treloar_photoelastic_1954,treloar_physics_1975,treloar_non-gaussian_1979,wu_improved_1992,wu_improved_1993,wu_network_1995,boyce_constitutive_2000} and has  been successfully used to describe the mechanics of semiflexible networks in a variety of contexts \cite{mackintosh_elasticity_1995,kroy_force-extension_1996,gardel_elastic_2004,storm_nonlinear_2005,broedersz_effective-medium_2009,lin_origins_2010,yao_elasticity_2010,cioroianu_normal_2013,van_oosterwyck_affine_2013,holzapfel_affine_2014,unterberger_advances_2014,meng_theory_2017,song_hyperelastic_2022}.

In this model, each filament segment is treated as a central-force elastic element of initial end-to-end distance $\ell_0$ with initial orientation $\hat{\mathbf{n}}_0$ that obeys a force-extension relation $\tau(\delta\ell)$, with $\delta\ell = \ell - \ell_0$ denoting the change in the deformed filament end-to-end distance $\ell$ with respect to the initial distance $\ell_0$. In the unstrained initial configuration, $\delta\ell = 0$ and $\tau = 0$.  The deformation gradient tensor $\bm{\mathrm{\Lambda}}$ transforms the initial filament end-to-end vector $\mathbf{r}_0 = \ell_0 \hat{\mathbf{n}}_0$ into the deformed end-to-end vector $\mathbf{r} = \bm{\Lambda} \mathbf{r}_0 = \ell \hat{\mathbf{n}}$ with a new orientation $\hat{\mathbf{n}}=\bm{\Lambda}\hat{\mathbf{n}}_0/|\bm{\Lambda}\hat{\mathbf{n}}_0|$ and end-to-end length $\ell = \ell_0 |\bm{\Lambda}\hat{\mathbf{n}}_0|$, such that the change in length is $\delta\ell = \ell_0( |\mathbf{\Lambda}\hat{\mathbf{n}}_0|- 1)$ with corresponding tension  $\bm{\tau}(\delta\ell)=\tau(\delta\ell)\hat{\mathbf{n}}$. We assume that the initial orientations $\hat{\mathbf{n}}_0$ are isotropically distributed and define the initial length density $\rho_0=\ell_0/V_0$, in which $V_0$ is the initial volume per filament segment. The length density in the deformed configuration becomes $\rho \equiv \langle \ell \rangle /V=  \rho_0 \langle  | \bm{\Lambda} \hat{\mathbf{n}}_0| \rangle /\det\bm{\Lambda}$. Averaging over all filaments, we  compute the Cauchy stress tensor $\bm{\sigma}  = V^{-1} \langle \bm{\tau}(\delta\ell)\otimes \mathbf{r} \rangle = \rho \langle \bm{\tau}(\delta\ell) \otimes \hat{\mathbf{n}} \rangle$ \cite{larson_structure_1999}, or equivalently  \cite{morse_viscoelasticity_1998-2,gittes_dynamic_1998,storm_nonlinear_2005, broedersz_modeling_2014}
\begin{equation} \label{eq:cauchystress}
\bm{\sigma} = \frac{\rho_0}{\rm{det}\mathbf{\Lambda}}\big\langle\bm{\tau}(\delta\ell) \otimes \left(\bm{\Lambda}\hat{\mathbf{n}}_0\right)\big\rangle,
\end{equation}
and the first Piola-Kirchhoff stress tensor $\mathbf{P}$ as
\begin{equation} \label{eq:pkstress}
\mathbf{P} = \det\mathbf{\Lambda} \bm{\sigma}^\mathrm{T} (\mathbf{\Lambda}^\mathrm{T})^{-1},
\end{equation}
the latter of which can be interpreted as the force in the deformed configuration per unit area in the undeformed configuration \cite{cioroianu_normal_2013}. Note that $\mathbf{P} = \bm{\sigma}$ in the small strain limit.

To quantify the alignment of filaments as a function of strain, we compute the symmetric and traceless nematic tensor $\bm{\mathrm{Q}}$ \cite{de_gennes_physics_2013,feng_alignment_2015} as 
\begin{equation} \label{eq:qtensor}
\bm{\mathrm{Q}}\equiv \bigg\langle \hat{\mathbf{n}}\otimes\hat{\mathbf{n}} -\frac{1}{d}\bm{\mathrm{I}}\bigg\rangle 
\end{equation}
in which $d = 3$ is the dimensionality and $\bm{\mathrm{I}}$ is the identity tensor. From the dominant eigenvalue $\lambda$ of $\bm{\mathrm{Q}}$, we obtain the nematic order parameter $S = (d/(d-1))\lambda$, which quantifies the filament alignment \cite{de_gennes_physics_2013}.

In the linear (small strain) regime, the nematic tensor $\mathbf{Q}$ and Cauchy stress tensor $\bm{\sigma}$ are related by the \textit{stress-optical law} \cite{doi_introduction_2004,macosko_rheology_1994},
\begin{equation} \label{eq:stress_optical_law}
    \mathbf{Q} = C \left( \bm{\sigma} - p \mathbf{I}\right)
\end{equation}
in which $C$ is a material-dependent proportionality coefficient and $p = \bm{\sigma} : \mathbf{I}$.

\begin{figure*}[ht!]
    \centering
    \includegraphics[width=1\textwidth]{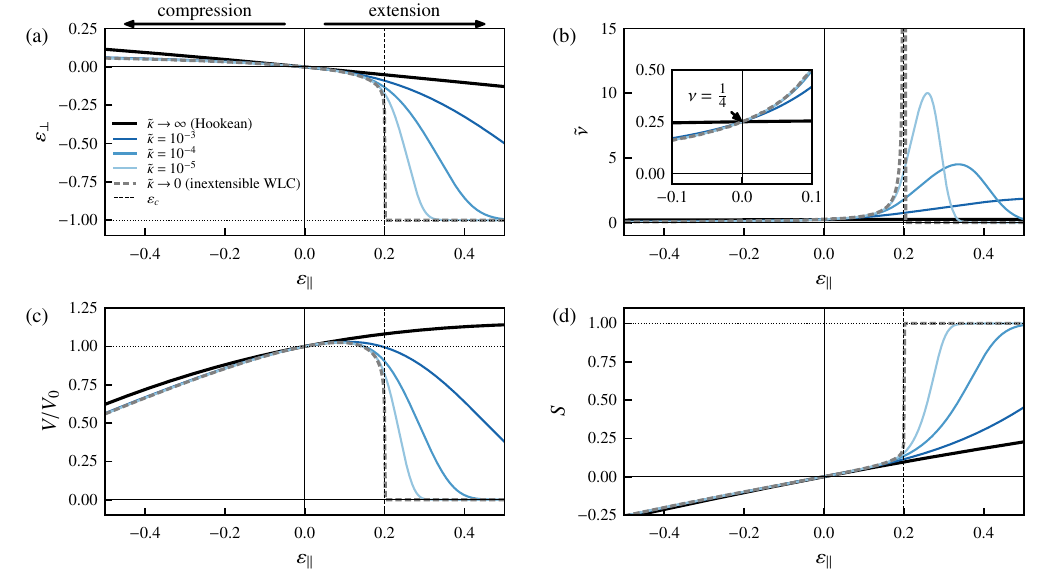}
    \caption{\label{varying_kappa} Response of initially isotropic networks of extensible wormlike chains with varying dimensionless bending rigidity $\tilde{\kappa} = \kappa/(\mu \ell_c^2)$ to applied uniaxial strain $\varepsilon_{\parallel}$ under the condition of zero transverse stress ($\sigma_\perp = 0$). Curves are also plotted for filaments in the Hookean limit ($\tilde{\kappa}\to \infty$ with finite $\mu$) and the inextensible WLC limit ( $\mu\to\infty$ with finite $\kappa$, such that $\tilde{\kappa}\to 0$). All curves correspond to $\ell_c = \ell_p = 1$.
        \textbf{(a)} Transverse strain $\varepsilon_\perp$ vs. applied longitudinal strain $\varepsilon_\parallel$.  In the limit of large $\kappa$, extensible WLC networks behave as Hookean spring networks, satisfying $\varepsilon_\perp=-\nu\varepsilon_\parallel$ up to relatively large strains. With decreasing $\kappa$, extensible wormlike chain networks exhibit increasingly rapid transverse contraction in the vicinity of the critical extensile strain $\varepsilon_c \equiv1/(6\tilde{\ell}_p - 1)$. In the inextensible WLC limit ($\kappa\to0$), the transverse collapse is complete ($\varepsilon_\perp \to -1)$ at $\varepsilon_\parallel = \varepsilon_c$. \textbf{(b)} The differential Poisson's ratio $\tilde{\nu} =-\partial\varepsilon_\perp/\partial\varepsilon_\parallel$ exhibits a peak that diverges in the limit of $\kappa\to0$ (or $\mu\to\infty$) at the critical extension $\varepsilon_c$. For small strains, $\tilde{\nu}  = \nu = 1/4$ for all models considered, as is expected for a Cauchy solid \cite{love_treatise_1906} \textbf{(c)} The relative volume $V/V_0$ initially increases with applied extension, before rapidly vanishing at the critical extension $\varepsilon_c$ in the $\kappa \to 0$ ($\mu \to \infty)$ limit. \textbf{(d)} The nematic alignment $S$ increases with applied extension before rapidly approaching $1$ (perfect alignment) at the critical extension $\varepsilon_c$ in the $\kappa \to 0$ ($\mu \to \infty$) limit. }
\end{figure*}

\subsection{Uniaxial strain protocol} \label{sec:strainprotocol}

Experimentally, Poisson's ratio can be obtained by stretching a material along one axis and measuring the resulting transverse strain \cite{love_treatise_1906}. To characterize the Poisson effect using the affine model, we apply a uniaxial strain $\varepsilon_{\parallel}$ along the $z$ axis and solve for the transverse strain $\varepsilon_\perp$ that satisfies a condition of zero stress along the transverse axes. The corresponding deformation gradient tensor $\bm{\Lambda}(\varepsilon_\parallel,\varepsilon_\perp)$ is
\begin{equation} \label{eq:deformationgradient}
\bm{\Lambda} = 
\begin{pmatrix}
1 + \varepsilon_\perp & 0 & 0 \\
0 & 1 + \varepsilon_\perp & 0\\ 
0 & 0 & 1 + \varepsilon_\parallel
 \end{pmatrix}.
\end{equation}
As the cross-sectional area varies considerably as a function of strain for these systems, the strain dependence of the first Piola-Kirchhoff stress tensor $\mathbf{P}$ (also called the engineering stress), given by Eq. \ref{eq:pkstress}, is more informative for our purposes than the Cauchy stress tensor $\bm{\sigma}$ (also called the true stress). This is because the Cauchy stress diverges as the cross-sectional area vanishes, even if the total tension remains finite. Note that $P_{ij} = 0$ if $\sigma_{ij} = 0$.   According to Eqs. \ref{eq:pkstress} and \ref{eq:deformationgradient}, the relevant components of $\mathbf{P}$ are given by 
\begin{equation}
P_\parallel = (1 + \varepsilon_\perp)^2 \sigma_\parallel
\end{equation}
and
\begin{equation}
P_\perp = (1 + \varepsilon_\perp)(1 + \varepsilon_\parallel)\sigma_\perp
\end{equation}
in which the components of $\bm{\sigma}$ are calculated using Eq. \ref{eq:cauchystress}.
Full details of the stress calculation are given in Appendix \ref{appendix:stress}. For a given applied $\varepsilon_\parallel$, we numerically solve Eq. \ref{eq:pkperp} for $\varepsilon_\perp(\varepsilon_\parallel)$ with $P_\perp(\varepsilon_\parallel, \varepsilon_\perp) = 0$. For small strains, $P_\parallel \approx E_0 \varepsilon_\parallel$, in which we have defined the linear Young's modulus $E_0 = \frac{1}{6} \rho_0 k \ell_0 / \ell_c$ (see Appendix \ref{appendix:stress}).

We can then calculate the differential Poisson's ratio $\tilde{\nu}$, defined as
\begin{equation} \label{eq:differential_poisson}
    \tilde{\nu} = \frac{\partial \varepsilon_\perp}{\partial\varepsilon_\parallel}
\end{equation}
and the relative volume,
\begin{equation} \label{eq:relative_volume}
    \frac{V}{V_0} = \frac{\rho_0}{\rho} = \det \bm{\Lambda}.
\end{equation}
In the limit of small strains,  $\lim_{\varepsilon\to 0^+}\tilde{\nu} = \nu = 1/4$ for this model and $V \approx V_0(1 + \varepsilon_\parallel/2)$.

We calculate the strain-dependent nematic alignment $S$  as \cite{de_gennes_physics_2013}
\begin{equation} \label{eq:nematic_alignment}
    S = \frac{3}{2} \bigg\langle \cos^2 \theta - \frac{1}{3}\bigg\rangle
\end{equation}
in which $\theta$ is the angle between the unit orientation vector of a given filament $\hat{\mathbf{n}}$ and the axis of applied strain ($\hat{\mathbf{z}}$) and the average is taken over all filaments. For small strains, $S \approx \varepsilon_\parallel/2$. See Appendix \ref{appendix:alignment} for additional details.

Since $P_\parallel \approx E_0 \varepsilon_\parallel$ and $S \approx \varepsilon_\parallel/2$ for small strains, the stress in this regime can be written as a function of the alignment, $P_\parallel \approx 2 E_0 S$. Consequently, the proportionality coefficient $C$ in the stress-optical law (Eq. \ref{eq:stress_optical_law}) is given by $C = 1/(2 E_0)$.

\begin{figure*}[htb!]
    \centering
    \includegraphics[width=1\textwidth]{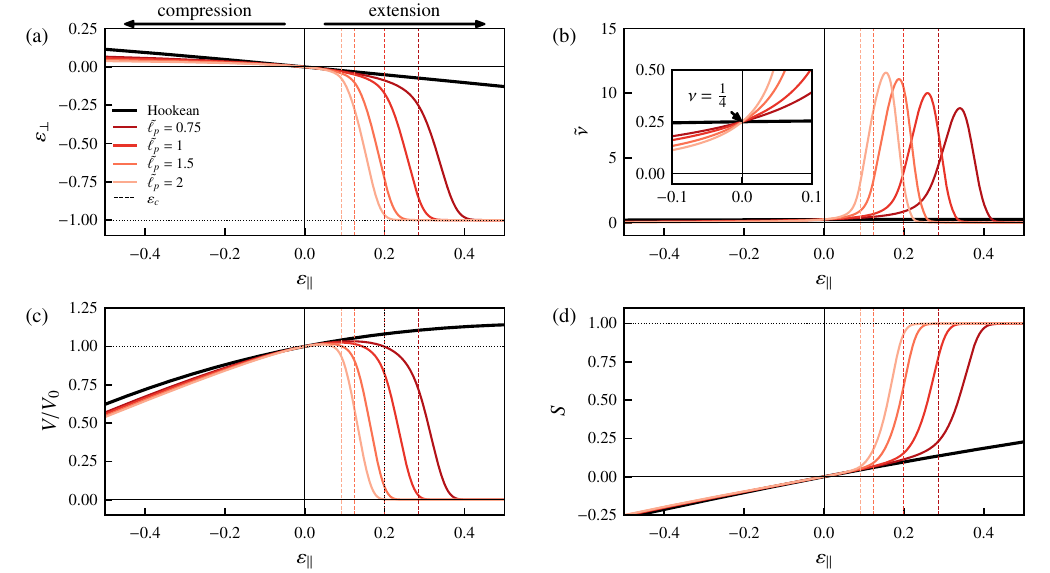}
    \caption{\label{varying_lp} Response of initially isotropic networks with varying dimensionless  persistence length $\tilde{\ell}_p \equiv \ell_p/\ell_c$ to applied uniaxial strain $\varepsilon_{\parallel}$ under the condition of zero transverse stress ($\sigma_\perp = 0$). For the plotted curves, the dimensionless bending rigidity is $\tilde{\kappa} = 10^{-5}$ except in the Hookean case($\tilde{\kappa}\to\infty$ with $\mu = 1$). Increasing the dimensionless persistence length $\tilde{\ell}_p$ leads to a decrease in the critical extensile strain $\varepsilon_c$, i.e. alignment and transverse contraction at lower values of applied uniaxial extensile strain. \textbf{(a)} Transverse strain $\varepsilon_\perp$ vs. applied uniaxial strain $\varepsilon_{\parallel}$. \textbf{(b)} Differential Poisson's ratio $\tilde{\nu}$. \textbf{(c)} Relative volume $V/V_0$. \textbf{(d)} Nematic alignment $S$. }
\end{figure*}

\section{Results and discussion} \label{sec:results}

Using the affine model, we now explore the effects of varying filament properties on the response of a network to uniaxial strain. Specifically, we will consider the effects of independently varying the dimensionless parameters defined in Sec. \ref{sec:forceextension}: the dimensionless bending rigidity $\tilde{\kappa}\equiv \kappa/(\mu \ell_c^2)$ and dimensionless persistence length $\tilde{\ell_p}=\ell_p/\ell_c$. Note that independently varying these quantities in experiments may be challenging, as both may depend on one or more of the same experimental variables, e.g. the polymer concentration \cite{mackintosh_elasticity_1995}. Our aim here is simply to provide an intuitive feel for how each of these dimensionless quantities influences the network-level response.

First, we consider the effects of varying the dimensionless bending rigidity $\tilde{\kappa}$ with a fixed value of the dimensionless persistence length $\tilde{\ell_p}=1$.  In Fig. \ref{varying_kappa}a, we plot the transverse strain $\varepsilon_\perp$ as a function of applied longitudinal strain $\varepsilon_\parallel$ for networks of extensible filaments over a wide range of $\tilde{\kappa}$ values, in addition to the Hookean spring limit ($\kappa \to \infty$ with finite $\mu$) and the inextensible WLC limit ($\mu \to \infty$ with finite $\kappa$). For small strains, all exhibit a linear regime in which $\varepsilon_\perp = -\nu \varepsilon_\parallel$, where the linear Poisson's ratio $\nu = 1/4$ is the expected value for a Cauchy solid \cite{love_treatise_1906, das_poissons_2010}. We find that as $\tilde{\kappa}$ increases, the behavior of the finite-$\tilde{\kappa}$ networks approaches that of simple Hookean spring networks, exhibiting an approximately linear dependence of transverse strain $\varepsilon_\perp$ on applied strain $\varepsilon_\parallel$ ( $\varepsilon_\perp \approx -\nu\varepsilon_\parallel$) over a large range of $\varepsilon_\parallel$. In contrast, as $\tilde{\kappa}$ is reduced, the behavior of the finite-$\tilde{\kappa}$ networks approaches that of inextensible WLC networks, exhibiting increasingly dramatic transverse contraction under finite values of applied strain $\varepsilon_\parallel$. Furthermore, we find that the value of applied strain $\varepsilon_\parallel$ corresponding to the inflection point in the $\varepsilon_\perp$ vs. $\varepsilon_\parallel$ curve decreases with decreasing $\tilde{\kappa}$. In the low-$\tilde{\kappa}$ limit, this inflection point approaches the critical extension $\varepsilon_c$  (Eq. \ref{eq:critical_extension}) corresponding to the applied extensional strain at which the end-to-end length of the most highly stretched filaments in the network (those oriented along the applied strain axis) approaches the filament contour length, $\delta\ell \to \ell_c^2/(6\ell_p)$. For inextensible WLC networks, complete transverse collapse occurs at $\varepsilon_c$.

This behavior is also reflected in the differential Poisson's ratio $\tilde{\nu} = -\partial \varepsilon_\perp/\partial \varepsilon_\parallel$ (Eq. \ref{eq:differential_poisson}), as we show in Fig. \ref{varying_kappa}b. The location of the inflection point in the $\varepsilon_\perp$ vs. $\varepsilon_\parallel$ curve corresponds to a peak in the differential Poisson's ratio $\tilde{\nu}$. This peak increases in magnitude and shifts to lower values of $\varepsilon_\parallel$ with decreasing $\tilde{\kappa}$, approaching $\varepsilon_c$ from above in the $\tilde{\kappa}\to 0$ limit.  The rapid transverse collapse that occurs at $\varepsilon_c$ in the low-$\kappa$ limit corresponds to a diverging differential Poisson's ratio $\tilde{\nu}$.  We can understand this behavior with a simple qualitative argument. Under finite applied strain, maintaining the zero transverse stress condition $\sigma_\perp=0$ requires a balance, in the transverse plane, between outward forces generated by compressed filaments and inward forces generated by stretched filaments. For networks of extensible wormlike chains (finite $\tilde{\kappa}$), increasing the applied extensional strain beyond the critical strain ($\varepsilon_\parallel > \varepsilon_c$) causes an increasing fraction of the stretched filaments to enter the enthalpic stretching regime ($\tau \sim \mu$), such that the inward transverse component of the resulting tension becomes proportional to $\mu$. However, the outward forces generated by compressed filaments remain proportional to $\kappa$ even in the large compression limit. For $\kappa\ll\mu$, achieving force balance thus requires significant transverse contraction as the elongated filaments enter the stretching-dominated regime.  In the inextensible WLC limit ($\mu\to\infty$) such a balance is impossible, as $\tau\to\infty$ as each filament approaches its contour length, so the network collapses completely ($\varepsilon_\perp\to-1$) precisely at the critical extensional strain $\varepsilon_c$.

An important consequence of the nonlinear Poisson effect is a reduction in the volume occupied by the network as it stiffens. Having determined $\varepsilon_{\perp}$ as a function of $\varepsilon_{\parallel}$, we compute the relative volume $V/V_0$ given by Eq. \ref{eq:relative_volume}, i.e. the ratio of the volume $V$ occupied by a portion of the strained network to its initial volume $V_0$.  In Fig. \ref{varying_kappa}c, we plot $V/V_0$ as a function of applied extension $\varepsilon_\parallel$ for the same systems as in the upper two panels. In the linear regime, $V/V_0\approx 1+\varepsilon_\parallel/2$. For a network of Hookean springs ($\tilde{\kappa} \to \infty$ with finite $\mu$), the relative volume $V/V_0$ decreases monotonically with compression and increases monotonically with extension over a large range of $\varepsilon_{\parallel}$. For networks with finite or zero $\tilde{\kappa}$ under applied compression, $V/V_0$ likewise decreases  and exhibits only a weak dependence on $\tilde{\kappa}$. Under applied extension, however, the relative volume initially increases in the linear regime before decreasing dramatically in the vicinity of the critical extension $\varepsilon_c$ as $\varepsilon_{\perp} \to -1$. Note that, because our model does not account for steric interaction between filaments, the relative volume vanishes completely in the case of highly nonlinear filaments under large extension. In reality, if we assume the filaments are negligibly compressible, then $V/V_0$ should not decrease below $\phi_0$, the initial volume fraction of filaments. 

The nonlinear Poisson effect is also characterized by the onset of filament alignment along the strain axis \cite{vader_strain-induced_2009}, quantified by the nematic alignment $S$ given by Eq. \ref{eq:nematic_alignment}. In the linear regime, $S$ increases with applied extensional strain $\varepsilon_\parallel$ as $S \approx \varepsilon_\parallel/2$. In the Hookean spring network limit ($\tilde{\kappa}\to\infty$ with finite $\mu$), the linear regime extends over a large range of extensile and compressive strain. For wormlike chain filaments with small or zero $\tilde{\kappa}$, identical behavior is seen in the linear regime, whereas we see rapidly increasing alignment in the vicinity of the critical extensional strain $\varepsilon_c$. In the $\tilde{\kappa} \to 0$ limit (inextensible WLCs), the network aligns completely ($S \to 1$) as $\varepsilon_\parallel \to \varepsilon_c$. This enhanced alignment is a direct consequence of the enhanced transverse contraction that occurs near $\varepsilon_{\parallel}\sim\varepsilon_c$ due to the asymmetric, nonlinear force-extension relationship of the individual filaments. Notably, significant alignment that coincides with stiffening and the nonlinear Poisson effect has been observed experimentally in uniaxially stretched fibrin \cite{vader_strain-induced_2009} and collagen gels \cite{ban_strong_2019}. Interestingly, under compression, we find that $S$ depends very little on $\tilde{\kappa}$ or, as we will soon see, $\tilde{\ell}_p$.

We now consider the effects of varying the dimensionless persistence length $\tilde{\ell}_p = \ell_p/\ell_c$ while the dimensionless bending rigidity $\tilde{\kappa}$ remains fixed. Recall, the critical extensional strain $\varepsilon_c$ (Eq. \ref{eq:critical_extension}) decreases as the dimensionless persistence length $\tilde{\ell}_p$ increases. Thus, increasing $\tilde{\ell_p}$ should cause the rapid transverse contraction, peak in the Poisson's ratio, and alignment to occur at lower values of applied extensional strain $\varepsilon_\parallel$.  This is precisely what is shown in Fig. \ref{varying_lp}, in which we plot the same quantities as in Fig. \ref{varying_kappa} for networks with fixed dimensionless bending rigidity ($\tilde{\kappa} = 10^{-5}$) and varying $\tilde{\ell}_p \in [0.5,2]$. Aside from the shift in $\varepsilon_c$ with varying $\tilde{\ell}_p$, the qualitative features (e.g. dramatic alignment and a large peak in the differential Poisson's ratio) remain unchanged.

\begin{figure}[htb!]
    \centering
    \includegraphics[width=1\columnwidth]{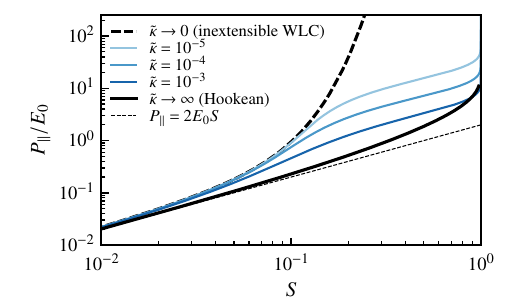}
    \caption{\label{fig:stress_vs_alignment} Filament alignment $S$ acts as a sensitive measure of the stress $P_\parallel$, and the stress-alignment relationship becomes increasingly nonlinear with decreasing bending rigidity $\tilde{\kappa}$. Here, we plot the longitudinal component of the first Piola-Kirchhoff stress, $P_\parallel$ normalized by the linear Young's modulus $E_0$, as a function of the nematic alignment $S$. The dashed line corresponds to the linear relationship valid for small strains, $P_\parallel = 2 E_0 S$.}
\end{figure}

We also examine the utility of the nematic alignment $S$ as an indicator of the stress $P_\parallel$. In the linear regime, the two quantities are directly proportional, with $P_\parallel \approx 2 E_0 S$ as we outline in Appendix \ref{appendix:alignment}. Thus, if the linear Young's modulus $E_0$ (Eq. \ref{eq:linear_youngsmod}) is known and the applied strain is sufficiently small, one can in principle use optical measurements of the nematic alignment $S$ as an indirect measure of the stress $P_\parallel$. This is simply another way of stating the ``stress optical law'' \cite{doi_introduction_2004,macosko_rheology_1994} (Eq. \ref{eq:stress_optical_law}), which relates the nematic and stress tensors for simple polymeric systems under small strains. For our systems, the proportionality coefficient in Eq. \ref{eq:stress_optical_law} is given by $C = 1/(2 E_0)$. As is shown in Fig. \ref{fig:stress_vs_alignment}, beyond the linear regime, this relationship no longer remains valid, and the deviation becomes more significant as $\tilde{\kappa}$ decreases. Nevertheless, if $E_0$, $\tilde{\kappa}$, and $\tilde{\ell}_p$ are known, one could in principle use a plot like Fig. \ref{fig:stress_vs_alignment} to estimate local values of the $P_\parallel$ from $S$ even beyond the linear regime.

\section{Conclusions} \label{sec:conclusion}

Here, we have provided a detailed characterization of the nonlinear Poisson effect in a simple affine model of semiflexible polymer networks over a broad range of parameters. The mechanical behavior of the model network is highly sensitive to the asymmetric force-extension relationship of the constituent filaments, which is governed by two key dimensionless parameters: the dimensionless bending rigidity $\tilde{\kappa}$ and the dimensionless persistence length $\tilde{\ell}_p$, which together control the force-extension relationship of individual chains. Microscopically, the dimensionless bending rigidity $\tilde{\kappa}$ essentially controls the asymmetry of the filament force-extension relationship, whereas the dimensionless persistence length $\tilde{\ell}_p$ controls the amount of initial contraction and thus the elongation required to straighten the filament. We find that, on the level of the full network, decreasing $\tilde{\kappa}$ intensifies the nonlinear Poisson effect and, in turn, intensifies the simultaneous alignment, densification, and stiffening of the network. The dimensionless persistence length $\tilde{\ell}_p$, in contrast, controls the amount of applied strain $\varepsilon_\parallel$ required to induce this effect: increasing $\tilde{\ell}_p$ reduces the critical extension $\varepsilon_c$, shifting the peak in the differential Poisson's ratio $\tilde{\nu}$ to lower values of $\varepsilon_\parallel$.

In future work, several extensions can be made to the model, both at the filament and network level, to improve its quantitative accuracy.  For example, alternative models for the force-extension relationship of semiflexible filaments that properly incorporate buckling \cite{odijk_microfibrillar_1998,emanuel_buckling_2007,baczynski_stretching_2008,blundell_buckling_2009,broedersz_modeling_2014,kurzthaler_elastic_2018} and more accurately account for filament extensibility \cite{unterberger_new_2013,holzapfel_elasticity_2013,unterberger_advances_2014} could be adopted in place of the relationship used here. We also note a recent model introducing a simplified force-extension relation based on pre-bent filaments, which permits simpler analytic results for some network properties \cite{li_fibrous_2022}.
On the network level, contributions to the total stress from the steric repulsion acting between filaments could also be included; this would prevent the complete transverse collapse of the network, with the initial filament volume fraction $\phi_0$ acting as a lower bound for the relative volume $V/V_0$, assuming incompressibility of the filaments. For hydrogels of collagen, fibrin or other biopolymers, the network volume reduction seen under extensional strain requires the expulsion of water, which will tend to be suppressed on short time scales \cite{janmey_negative_2007, de_cagny_porosity_2016,vahabi_normal_2018}. Thus, our results here are limited to the long-time elastic regime of network response. In future work, it would be interesting to consider effects of both network and solvent \cite{doi_soft_2013,yamamoto_large_2017} and poroelasticity \cite{de_cagny_porosity_2016,vahabi_normal_2018, punter_poroelasticity_2020}, which would both improve accuracy of the model and allow for consideration of the dynamics associated with the nonlinear Poisson effect.

We note that, for the affine network model, the critical extensional strain $\varepsilon_c$ corresponding to the nonlinear Poisson effect can be related to the critical shear strain $\gamma_c$ corresponding to strain stiffening under applied simple shear. The maximum extension $\lambda_\mathrm{max}$ for a given deformation $\mathbf{\Lambda}$ can be determined from the maximum eigenvalue $\lambda_\mathrm{max}^2$ of the right Cauchy-Green tensor $\mathbf{C} = \mathbf{\Lambda}^T\mathbf{\Lambda}$. For $\mathbf{\Lambda}(\gamma)$ corresponding to simple shear $\gamma$, $\lambda_\mathrm{max}^2=1+(\gamma^2+\gamma\sqrt{\gamma^2+4})/2$, and equivalently $\gamma=\lambda_\mathrm{max}-1/\lambda_\mathrm{max}$. As strain-stiffening occurs when $\lambda_\mathrm{max} = 1+\varepsilon_c$, we can therefore relate the critical shear strain to the critical extension corresponding to the nonlinear Poisson's ratio as $\gamma_c = 1+\varepsilon_c - 1/(1+\varepsilon_c)$. 

While an affine model such as the one we propose here can be a useful, and in some cases, even predictive approach to modeling networks, we note that non-affine deformations are known to be important in some cases \cite{kroy_force-extension_1996,head_deformation_2003,wilhelm_elasticity_2003,heussinger_floppy_2006,broedersz_modeling_2014}. When considering non-affine effects, it is important to include the effects of mechanical integrity and bending rigidity of semiflexible filaments across network nodes, which leads to possible long-range mechanical effects \cite{broedersz_filament-length-controlled_2012}. This makes the modeling of such networks substantially more challenging. Nevertheless, it may be possible to extend a recent method for modeling such effects on linear elasticity to the nonlinear regime \cite{chen_nonaffine_2023}. We speculate that non-affine effects can result in a possible delayed onset of network nonlinearity \cite{gardel_elastic_2004} and a possible reduction of the the peak in the Poisson ratio we see above.

\acknowledgments{We thank Michael Rubinstein and Sergey Panyukov for stimulating discussions. This work was supported in part by the National Science Foundation Division of Materials Research (Grant No. DMR-2224030) and the National Science Foundation Center for Theoretical Biological Physics (Grant No. PHY-2019745). J.L.S. is grateful for the support of the Eric and Wendy Schmidt AI in Science Postdoctoral Fellowship at the University of Chicago.  We would also like to thank the Isaac Newton Institute for Mathematical Sciences, Cambridge, for support and hospitality during the programme ``New statistical physics in living matter: non equilibrium states under adaptive control,'' where work on this paper was undertaken. This work was supported by EPSRC grant no. EP/R014604/1.
}

\appendix

\section{Calculation of the stress components} \label{appendix:stress}

According to the affine network model (Eq. \ref{eq:cauchystress}), the $ij$ component of the Cauchy stress tensor $\bm{\sigma}$ for an initially isotropic network can be written as \begin{equation} \label{stresstensor}
\sigma_{ij}=\frac{\rho_0}{4\pi\mathrm{det}\bm{\Lambda}}\iint d\theta_0 d\varphi_0 \sin\theta_0\left[\tau(\delta\ell)\frac{(\bm{\Lambda}\hat{\mathbf{n}}_0)_i (\bm{\Lambda}\hat{\mathbf{n}}_0)_j}{\left|\bm{\Lambda}\hat{\mathbf{n}}_0\right|}\right].
\end{equation}
in which $\rho_0$ is the initial length density, $\bm{\Lambda}$ is the deformation gradient tensor, $\tau(\delta\ell)$ is the force-extension relationship, $\hat{\mathbf{n}}_0(\theta_0,\phi_0)$ is the initial filament orientation defined as
\begin{equation} \label{eq:initialorientation}
    \hat{\mathbf{n}}_0(\theta_0,\varphi_0) = \begin{pmatrix}\sin\theta_0\cos\varphi_0 \\ \sin\theta_0\sin\varphi_0 \\ \cos\theta_0 \end{pmatrix},
\end{equation}
and the integrals are taken over the ranges $0 \le \theta_0 \le \pi$ and $0\le \varphi_0 \le 2\pi$.

For the uniaxial strain scenario considered in this work, the appropriate deformation gradient tensor $\bm{\Lambda}(\varepsilon_\parallel, \varepsilon_\perp)$ is given by Eq. \ref{eq:deformationgradient}, for which $\mathrm{det}\bm{\Lambda} = (1+\varepsilon_\perp)^2(1+\varepsilon_\parallel)$.
Taking advantage of the axial symmetry of the system, we can write the normal components of the first Piola-Kirchhoff stress tensor as

\begin{align} \label{eq:pkparallel}
    P_{\parallel} = \frac{\rho_0(1+\varepsilon_\parallel)}{2}\int d\theta_0\sin\theta_0 \left[\tau(\delta\ell)\frac{\cos^2 \theta_0}{|\bm{\Lambda}\hat{\mathbf{n}}_0|}\right]
\end{align}
\begin{align} \label{eq:pkperp}
    P_{\perp} = \frac{\rho_0(1+\varepsilon_\perp)}{4}\int d\theta_0\sin\theta_0\left[\tau(\delta\ell)\frac{\sin^2 \theta_0 }{|\bm{\Lambda}\hat{\mathbf{n}}_0|}\right]
\end{align}
in which $|\bm{\Lambda}\hat{\mathbf{n}}_0| = \sqrt{(1+\varepsilon_\perp)^2\sin^2\theta_0 + (1+\varepsilon_\parallel)^2\cos^2\theta_0}$. As described in the main text, the change in length is $\delta\ell = \ell_0\left(|\bm{\Lambda}\hat{\mathbf{n}}_0|-1\right)$ and the tension $\tau(\delta\ell)$ is determined by inversion of Eq. \ref{eq:force_extension}. 

In the small strain regime, we can solve Eqs. \ref{eq:pkparallel} and \ref{eq:pkperp} using the linear force-extension relationship $\tau = k\delta\ell/\ell_c$, with $k = 1/(\mu^{-1} + \ell_c^3/(90\kappa\ell_p))$, the linear strain dependence $\varepsilon_\perp = -\nu \varepsilon_\parallel$ with $\nu = 1/4$. To linear order in $\varepsilon_\parallel$, the parallel stress component $P_\parallel$ behaves as
\begin{equation}
    P_\parallel \approx \sigma_\parallel \approx \frac{1}{6}\rho_0 k\frac{\ell_0}{\ell_c}\varepsilon_\parallel
\end{equation}
from which we see that the linear Young's modulus $E_0=\lim_{\varepsilon\to0^+}\sigma_\parallel/\varepsilon_\parallel$ is
\begin{equation} \label{eq:linear_youngsmod}
    E_0 = \frac{1}{6} \rho_0k\frac{\ell_0}{\ell_c}. 
\end{equation}

\section{Calculation of the nematic alignment} \label{appendix:alignment}

As described in the main text, we consider an initially isotropic network with orientations $\hat{\mathbf{n}}_0$ given by Eq. \ref{eq:initialorientation} under deformation gradient tensor $\bm{\Lambda}$, such that the filament orientations in the deformed configuration are $\hat{\mathbf{n}} = \bm{\Lambda}\hat{\mathbf{n}}_0/|\bm{\Lambda}\hat{\mathbf{n}}_0|$. In the deformed configuration, the traceless and symmetric nematic tensor $\mathbf{Q}$ is given by Eq. \ref{eq:qtensor} and can be written in terms of its eigenvalues $\lambda_i$ and orthonormal eigenvectors $\hat{\mathbf{v}}_i$ as 
\begin{equation}
    \mathbf{Q} = \lambda_1 \hat{\mathbf{v}}_1\otimes\hat{\mathbf{v}}_1 + \lambda_2 \hat{\mathbf{v}}_2 \otimes \hat{\mathbf{v}}_2 + \lambda_3 \hat{\mathbf{v}}_3 \otimes \hat{\mathbf{v}}_3.
\end{equation}
with $\lambda_1+\lambda_2+\lambda_3 = 0$. For uniaxial deformation applied along the $z$-axis, symmetry requires that two of the eigenvalues of $\mathbf{Q}$ are equal, so we can write $\lambda_1 = \lambda_2 = -S/3$ and $\lambda_3 = 2S/3$ \cite{ball_mathematics_2012}. Noting that $\hat{\mathbf{v}}_3 = \hat{\mathbf{z}}$ in this case, we can write the tensor $\mathbf{Q}$ as
\begin{equation}
    \mathbf{Q} = S\left(\hat{\mathbf{z}}\otimes\hat{\mathbf{z}}-\frac{1}{3}\mathbf{I}\right)
\end{equation}
where the nematic alignment $S$ is given by Eq. \ref{eq:nematic_alignment}. Integrating over the initial orientations $\hat{\mathbf{n}}_0$, we find
\begin{align}
    S &= \frac{1}{4\pi}\iint d\varphi_0 d\theta_0 \sin\theta_0 \left[\frac{3}{2}\left(\frac{(1+\varepsilon_\parallel)^2\cos^2\theta_0}{|\bm{\Lambda}\hat{\mathbf{n}}_0|^2}-\frac{1}{3}\right)\right] \nonumber \\
    &= \frac{1}{2}\int d\theta_0 \sin\theta_0 \left[\frac{3}{2}\left(\frac{(1+\varepsilon_\parallel)^2\cos^2\theta_0}{|\bm{\Lambda}\hat{\mathbf{n}}_0|^2}-\frac{1}{3}\right)\right] \nonumber \\   & =\frac{3}{2}\frac{(1+\varepsilon_\parallel)^2}{a^2}\left( 1-\left(\frac{1+\varepsilon_\perp}{a}\right)\tan^{-1}\left(\frac{a}{1+\varepsilon_\perp}\right)\right)-\frac{1}{2}                    
\end{align}
in which $a=\sqrt{(1+\varepsilon_\parallel)^2-(1+\varepsilon_\perp)^2}$. For small $\varepsilon_\parallel$, this yields $S \approx \varepsilon_\parallel/2$.
Note that the alignment $S$ can vary in the range $S\in[-1/2,1]$, in which $S = -1/2$ corresponds to a configuration in which all filaments are oriented perpendicular to $\hat{\mathbf{z}}$ and $S=1$ corresponds to a configuration in which all filaments are parallel to $\hat{\mathbf{z}}$.

In the small strain regime, in which $\varepsilon_\perp = -\nu \varepsilon_\parallel$ with $\nu = 1/4$, we have
\begin{equation}
    S \approx \frac{1}{2}\varepsilon_\parallel
\end{equation}
Since $P_\parallel \approx E_0 \varepsilon_\parallel$ in this regime, we find that the relationship between stress and alignment in the linear regime is
\begin{equation} \label{eq:stress_alignment}
P_\parallel \approx 2 E_0 S
\end{equation}
with $E_0$ given by Eq. \ref{eq:linear_youngsmod}. This relationship is plotted in Fig. \ref{fig:stress_vs_alignment}.
As we noted in the main text, Eq. \ref{eq:stress_alignment} implies that the proportionality coefficient $C$ in Eq. \ref{eq:stress_optical_law} is given by $C = 1/(2 E_0)$.

\bibliography{Bibliography}{}

\begin{thebibliography}{89}%
\makeatletter
\providecommand \@ifxundefined [1]{%
 \@ifx{#1\undefined}
}%
\providecommand \@ifnum [1]{%
 \ifnum #1\expandafter \@firstoftwo
 \else \expandafter \@secondoftwo
 \fi
}%
\providecommand \@ifx [1]{%
 \ifx #1\expandafter \@firstoftwo
 \else \expandafter \@secondoftwo
 \fi
}%
\providecommand \natexlab [1]{#1}%
\providecommand \enquote  [1]{``#1''}%
\providecommand \bibnamefont  [1]{#1}%
\providecommand \bibfnamefont [1]{#1}%
\providecommand \citenamefont [1]{#1}%
\providecommand \href@noop [0]{\@secondoftwo}%
\providecommand \href [0]{\begingroup \@sanitize@url \@href}%
\providecommand \@href[1]{\@@startlink{#1}\@@href}%
\providecommand \@@href[1]{\endgroup#1\@@endlink}%
\providecommand \@sanitize@url [0]{\catcode `\\12\catcode `\$12\catcode `\&12\catcode `\#12\catcode `\^12\catcode `\_12\catcode `\%12\relax}%
\providecommand \@@startlink[1]{}%
\providecommand \@@endlink[0]{}%
\providecommand \url  [0]{\begingroup\@sanitize@url \@url }%
\providecommand \@url [1]{\endgroup\@href {#1}{\urlprefix }}%
\providecommand \urlprefix  [0]{URL }%
\providecommand \Eprint [0]{\href }%
\providecommand \doibase [0]{https://doi.org/}%
\providecommand \selectlanguage [0]{\@gobble}%
\providecommand \bibinfo  [0]{\@secondoftwo}%
\providecommand \bibfield  [0]{\@secondoftwo}%
\providecommand \translation [1]{[#1]}%
\providecommand \BibitemOpen [0]{}%
\providecommand \bibitemStop [0]{}%
\providecommand \bibitemNoStop [0]{.\EOS\space}%
\providecommand \EOS [0]{\spacefactor3000\relax}%
\providecommand \BibitemShut  [1]{\csname bibitem#1\endcsname}%
\let\auto@bib@innerbib\@empty
\bibitem [{\citenamefont {Kasza}\ \emph {et~al.}(2007)\citenamefont {Kasza}, \citenamefont {Rowat}, \citenamefont {Liu}, \citenamefont {Angelini}, \citenamefont {Brangwynne}, \citenamefont {Koenderink},\ and\ \citenamefont {Weitz}}]{kasza_cell_2007}%
  \BibitemOpen
  \bibfield  {author} {\bibinfo {author} {\bibfnamefont {K.~E.}\ \bibnamefont {Kasza}}, \bibinfo {author} {\bibfnamefont {A.~C.}\ \bibnamefont {Rowat}}, \bibinfo {author} {\bibfnamefont {J.}~\bibnamefont {Liu}}, \bibinfo {author} {\bibfnamefont {T.~E.}\ \bibnamefont {Angelini}}, \bibinfo {author} {\bibfnamefont {C.~P.}\ \bibnamefont {Brangwynne}}, \bibinfo {author} {\bibfnamefont {G.~H.}\ \bibnamefont {Koenderink}},\ and\ \bibinfo {author} {\bibfnamefont {D.~A.}\ \bibnamefont {Weitz}},\ }\bibfield  {title} {\bibinfo {title} {The cell as a material},\ }\href {https://doi.org/10.1016/j.ceb.2006.12.002} {\bibfield  {journal} {\bibinfo  {journal} {Current Opinion in Cell Biology}\ }\textbf {\bibinfo {volume} {19}},\ \bibinfo {pages} {101} (\bibinfo {year} {2007})}\BibitemShut {NoStop}%
\bibitem [{\citenamefont {Pritchard}\ \emph {et~al.}(2014)\citenamefont {Pritchard}, \citenamefont {Shery~Huang},\ and\ \citenamefont {Terentjev}}]{pritchard_mechanics_2014}%
  \BibitemOpen
  \bibfield  {author} {\bibinfo {author} {\bibfnamefont {R.~H.}\ \bibnamefont {Pritchard}}, \bibinfo {author} {\bibfnamefont {Y.~Y.}\ \bibnamefont {Shery~Huang}},\ and\ \bibinfo {author} {\bibfnamefont {E.~M.}\ \bibnamefont {Terentjev}},\ }\bibfield  {title} {\bibinfo {title} {Mechanics of biological networks: From the cell cytoskeleton to connective tissue},\ }\href {https://doi.org/10.1039/c3sm52769g} {\bibfield  {journal} {\bibinfo  {journal} {Soft Matter}\ }\textbf {\bibinfo {volume} {10}},\ \bibinfo {pages} {1864} (\bibinfo {year} {2014})}\BibitemShut {NoStop}%
\bibitem [{\citenamefont {Broedersz}\ and\ \citenamefont {MacKintosh}(2014)}]{broedersz_modeling_2014}%
  \BibitemOpen
  \bibfield  {author} {\bibinfo {author} {\bibfnamefont {C.~P.}\ \bibnamefont {Broedersz}}\ and\ \bibinfo {author} {\bibfnamefont {F.~C.}\ \bibnamefont {MacKintosh}},\ }\bibfield  {title} {\bibinfo {title} {Modeling semiflexible polymer networks},\ }\href {https://doi.org/10.1103/revmodphys.86.995} {\bibfield  {journal} {\bibinfo  {journal} {Reviews of Modern Physics}\ }\textbf {\bibinfo {volume} {86}},\ \bibinfo {pages} {995} (\bibinfo {year} {2014})}\BibitemShut {NoStop}%
\bibitem [{\citenamefont {Burla}\ \emph {et~al.}(2019)\citenamefont {Burla}, \citenamefont {Mulla}, \citenamefont {Vos}, \citenamefont {{Aufderhorst-Roberts}},\ and\ \citenamefont {Koenderink}}]{burla_mechanical_2019}%
  \BibitemOpen
  \bibfield  {author} {\bibinfo {author} {\bibfnamefont {F.}~\bibnamefont {Burla}}, \bibinfo {author} {\bibfnamefont {Y.}~\bibnamefont {Mulla}}, \bibinfo {author} {\bibfnamefont {B.~E.}\ \bibnamefont {Vos}}, \bibinfo {author} {\bibfnamefont {A.}~\bibnamefont {{Aufderhorst-Roberts}}},\ and\ \bibinfo {author} {\bibfnamefont {G.~H.}\ \bibnamefont {Koenderink}},\ }\bibfield  {title} {\bibinfo {title} {From mechanical resilience to active material properties in biopolymer networks},\ }\href {https://doi.org/10.1038/s42254-019-0036-4} {\bibfield  {journal} {\bibinfo  {journal} {Nature Reviews Physics}\ }\textbf {\bibinfo {volume} {1}},\ \bibinfo {pages} {249} (\bibinfo {year} {2019})}\BibitemShut {NoStop}%
\bibitem [{\citenamefont {Fung}(1967)}]{fung_elasticity_1967}%
  \BibitemOpen
  \bibfield  {author} {\bibinfo {author} {\bibfnamefont {Y.~C.}\ \bibnamefont {Fung}},\ }\bibfield  {title} {\bibinfo {title} {Elasticity of soft tissues in simple elongation},\ }\href {https://doi.org/10.1152/ajplegacy.1967.213.6.1532} {\bibfield  {journal} {\bibinfo  {journal} {American Journal of Physiology}\ }\textbf {\bibinfo {volume} {213}},\ \bibinfo {pages} {1532} (\bibinfo {year} {1967})}\BibitemShut {NoStop}%
\bibitem [{\citenamefont {Gardel}\ \emph {et~al.}(2004)\citenamefont {Gardel}, \citenamefont {Shin}, \citenamefont {MacKintosh}, \citenamefont {Mahadevan}, \citenamefont {Matsudaira},\ and\ \citenamefont {Weitz}}]{gardel_elastic_2004}%
  \BibitemOpen
  \bibfield  {author} {\bibinfo {author} {\bibfnamefont {M.~L.}\ \bibnamefont {Gardel}}, \bibinfo {author} {\bibfnamefont {J.~H.}\ \bibnamefont {Shin}}, \bibinfo {author} {\bibfnamefont {F.~C.}\ \bibnamefont {MacKintosh}}, \bibinfo {author} {\bibfnamefont {L.}~\bibnamefont {Mahadevan}}, \bibinfo {author} {\bibfnamefont {P.}~\bibnamefont {Matsudaira}},\ and\ \bibinfo {author} {\bibfnamefont {D.~A.}\ \bibnamefont {Weitz}},\ }\bibfield  {title} {\bibinfo {title} {Elastic {{Behavior}} of {{Cross-Linked}} and {{Bundled Actin Networks}}},\ }\href {https://doi.org/10.1126/science.1095087} {\bibfield  {journal} {\bibinfo  {journal} {Science}\ }\textbf {\bibinfo {volume} {304}},\ \bibinfo {pages} {1301} (\bibinfo {year} {2004})}\BibitemShut {NoStop}%
\bibitem [{\citenamefont {Storm}\ \emph {et~al.}(2005)\citenamefont {Storm}, \citenamefont {Pastore}, \citenamefont {MacKintosh}, \citenamefont {Lubensky},\ and\ \citenamefont {Janmey}}]{storm_nonlinear_2005}%
  \BibitemOpen
  \bibfield  {author} {\bibinfo {author} {\bibfnamefont {C.}~\bibnamefont {Storm}}, \bibinfo {author} {\bibfnamefont {J.~J.}\ \bibnamefont {Pastore}}, \bibinfo {author} {\bibfnamefont {F.~C.}\ \bibnamefont {MacKintosh}}, \bibinfo {author} {\bibfnamefont {T.~C.}\ \bibnamefont {Lubensky}},\ and\ \bibinfo {author} {\bibfnamefont {P.~A.}\ \bibnamefont {Janmey}},\ }\bibfield  {title} {\bibinfo {title} {Nonlinear elasticity in biological gels},\ }\href {https://doi.org/10.1038/nature03521} {\bibfield  {journal} {\bibinfo  {journal} {Nature}\ }\textbf {\bibinfo {volume} {435}},\ \bibinfo {pages} {191} (\bibinfo {year} {2005})}\BibitemShut {NoStop}%
\bibitem [{\citenamefont {Van~Oosten}\ \emph {et~al.}(2016)\citenamefont {Van~Oosten}, \citenamefont {Vahabi}, \citenamefont {Licup}, \citenamefont {Sharma}, \citenamefont {Galie}, \citenamefont {MacKintosh},\ and\ \citenamefont {Janmey}}]{van_oosten_uncoupling_2016}%
  \BibitemOpen
  \bibfield  {author} {\bibinfo {author} {\bibfnamefont {A.~S.~G.}\ \bibnamefont {Van~Oosten}}, \bibinfo {author} {\bibfnamefont {M.}~\bibnamefont {Vahabi}}, \bibinfo {author} {\bibfnamefont {A.~J.}\ \bibnamefont {Licup}}, \bibinfo {author} {\bibfnamefont {A.}~\bibnamefont {Sharma}}, \bibinfo {author} {\bibfnamefont {P.~A.}\ \bibnamefont {Galie}}, \bibinfo {author} {\bibfnamefont {F.~C.}\ \bibnamefont {MacKintosh}},\ and\ \bibinfo {author} {\bibfnamefont {P.~A.}\ \bibnamefont {Janmey}},\ }\bibfield  {title} {\bibinfo {title} {Uncoupling shear and uniaxial elastic moduli of semiflexible biopolymer networks: Compression-softening and stretch-stiffening},\ }\href {https://doi.org/10.1038/srep19270} {\bibfield  {journal} {\bibinfo  {journal} {Scientific Reports}\ }\textbf {\bibinfo {volume} {6}},\ \bibinfo {pages} {19270} (\bibinfo {year} {2016})}\BibitemShut {NoStop}%
\bibitem [{\citenamefont {Xu}\ and\ \citenamefont {Safran}(2017)}]{xu_compressive_2017}%
  \BibitemOpen
  \bibfield  {author} {\bibinfo {author} {\bibfnamefont {X.}~\bibnamefont {Xu}}\ and\ \bibinfo {author} {\bibfnamefont {S.~A.}\ \bibnamefont {Safran}},\ }\bibfield  {title} {\bibinfo {title} {Compressive elasticity of polydisperse biopolymer gels},\ }\href {https://doi.org/10.1103/PhysRevE.95.052415} {\bibfield  {journal} {\bibinfo  {journal} {Physical Review E}\ }\textbf {\bibinfo {volume} {95}},\ \bibinfo {pages} {052415} (\bibinfo {year} {2017})}\BibitemShut {NoStop}%
\bibitem [{\citenamefont {Vahabi}\ \emph {et~al.}(2016)\citenamefont {Vahabi}, \citenamefont {Sharma}, \citenamefont {Licup}, \citenamefont {Van~Oosten}, \citenamefont {Galie}, \citenamefont {Janmey},\ and\ \citenamefont {MacKintosh}}]{vahabi_elasticity_2016}%
  \BibitemOpen
  \bibfield  {author} {\bibinfo {author} {\bibfnamefont {M.}~\bibnamefont {Vahabi}}, \bibinfo {author} {\bibfnamefont {A.}~\bibnamefont {Sharma}}, \bibinfo {author} {\bibfnamefont {A.~J.}\ \bibnamefont {Licup}}, \bibinfo {author} {\bibfnamefont {A.~S.~G.}\ \bibnamefont {Van~Oosten}}, \bibinfo {author} {\bibfnamefont {P.~A.}\ \bibnamefont {Galie}}, \bibinfo {author} {\bibfnamefont {P.~A.}\ \bibnamefont {Janmey}},\ and\ \bibinfo {author} {\bibfnamefont {F.~C.}\ \bibnamefont {MacKintosh}},\ }\bibfield  {title} {\bibinfo {title} {Elasticity of fibrous networks under uniaxial prestress},\ }\href {https://doi.org/10.1039/C6SM00606J} {\bibfield  {journal} {\bibinfo  {journal} {Soft Matter}\ }\textbf {\bibinfo {volume} {12}},\ \bibinfo {pages} {5050} (\bibinfo {year} {2016})}\BibitemShut {NoStop}%
\bibitem [{\citenamefont {Van~Oosten}\ \emph {et~al.}(2019)\citenamefont {Van~Oosten}, \citenamefont {Chen}, \citenamefont {Chin}, \citenamefont {Cruz}, \citenamefont {Patteson}, \citenamefont {Pogoda}, \citenamefont {Shenoy},\ and\ \citenamefont {Janmey}}]{van_oosten_emergence_2019}%
  \BibitemOpen
  \bibfield  {author} {\bibinfo {author} {\bibfnamefont {A.~S.~G.}\ \bibnamefont {Van~Oosten}}, \bibinfo {author} {\bibfnamefont {X.}~\bibnamefont {Chen}}, \bibinfo {author} {\bibfnamefont {L.}~\bibnamefont {Chin}}, \bibinfo {author} {\bibfnamefont {K.}~\bibnamefont {Cruz}}, \bibinfo {author} {\bibfnamefont {A.~E.}\ \bibnamefont {Patteson}}, \bibinfo {author} {\bibfnamefont {K.}~\bibnamefont {Pogoda}}, \bibinfo {author} {\bibfnamefont {V.~B.}\ \bibnamefont {Shenoy}},\ and\ \bibinfo {author} {\bibfnamefont {P.~A.}\ \bibnamefont {Janmey}},\ }\bibfield  {title} {\bibinfo {title} {Emergence of tissue-like mechanics from fibrous networks confined by close-packed cells},\ }\href {https://doi.org/10.1038/s41586-019-1516-5} {\bibfield  {journal} {\bibinfo  {journal} {Nature}\ }\textbf {\bibinfo {volume} {573}},\ \bibinfo {pages} {96} (\bibinfo {year} {2019})}\BibitemShut {NoStop}%
\bibitem [{\citenamefont {{Ed-Daoui}}\ and\ \citenamefont {Snabre}(2021)}]{ed-daoui_poroviscoelasticity_2021}%
  \BibitemOpen
  \bibfield  {author} {\bibinfo {author} {\bibfnamefont {A.}~\bibnamefont {{Ed-Daoui}}}\ and\ \bibinfo {author} {\bibfnamefont {P.}~\bibnamefont {Snabre}},\ }\bibfield  {title} {\bibinfo {title} {Poroviscoelasticity and compression-softening of agarose hydrogels},\ }\href {https://doi.org/10.1007/s00397-021-01267-3} {\bibfield  {journal} {\bibinfo  {journal} {Rheologica Acta}\ }\textbf {\bibinfo {volume} {60}},\ \bibinfo {pages} {327} (\bibinfo {year} {2021})}\BibitemShut {NoStop}%
\bibitem [{\citenamefont {Brown}\ \emph {et~al.}(2009)\citenamefont {Brown}, \citenamefont {Litvinov}, \citenamefont {Discher}, \citenamefont {Purohit},\ and\ \citenamefont {Weisel}}]{brown_multiscale_2009}%
  \BibitemOpen
  \bibfield  {author} {\bibinfo {author} {\bibfnamefont {A.~E.~X.}\ \bibnamefont {Brown}}, \bibinfo {author} {\bibfnamefont {R.~I.}\ \bibnamefont {Litvinov}}, \bibinfo {author} {\bibfnamefont {D.~E.}\ \bibnamefont {Discher}}, \bibinfo {author} {\bibfnamefont {P.~K.}\ \bibnamefont {Purohit}},\ and\ \bibinfo {author} {\bibfnamefont {J.~W.}\ \bibnamefont {Weisel}},\ }\bibfield  {title} {\bibinfo {title} {Multiscale mechanics of fibrin polymer: {{Gel}} stretching with protein unfolding and loss of water},\ }\href {https://doi.org/10.1126/science.1172484} {\bibfield  {journal} {\bibinfo  {journal} {Science}\ }\textbf {\bibinfo {volume} {325}},\ \bibinfo {pages} {741} (\bibinfo {year} {2009})}\BibitemShut {NoStop}%
\bibitem [{\citenamefont {Vader}\ \emph {et~al.}(2009)\citenamefont {Vader}, \citenamefont {Kabla}, \citenamefont {Weitz},\ and\ \citenamefont {Mahadevan}}]{vader_strain-induced_2009}%
  \BibitemOpen
  \bibfield  {author} {\bibinfo {author} {\bibfnamefont {D.}~\bibnamefont {Vader}}, \bibinfo {author} {\bibfnamefont {A.}~\bibnamefont {Kabla}}, \bibinfo {author} {\bibfnamefont {D.}~\bibnamefont {Weitz}},\ and\ \bibinfo {author} {\bibfnamefont {L.}~\bibnamefont {Mahadevan}},\ }\bibfield  {title} {\bibinfo {title} {Strain-{{Induced Alignment}} in {{Collagen Gels}}},\ }\href {https://doi.org/10.1371/journal.pone.0005902} {\bibfield  {journal} {\bibinfo  {journal} {PLoS ONE}\ }\textbf {\bibinfo {volume} {4}},\ \bibinfo {pages} {e5902} (\bibinfo {year} {2009})}\BibitemShut {NoStop}%
\bibitem [{\citenamefont {Steinwachs}\ \emph {et~al.}(2016)\citenamefont {Steinwachs}, \citenamefont {Metzner}, \citenamefont {Skodzek}, \citenamefont {Lang}, \citenamefont {Thievessen}, \citenamefont {Mark}, \citenamefont {M{\"u}nster}, \citenamefont {Aifantis},\ and\ \citenamefont {Fabry}}]{steinwachs_three-dimensional_2016}%
  \BibitemOpen
  \bibfield  {author} {\bibinfo {author} {\bibfnamefont {J.}~\bibnamefont {Steinwachs}}, \bibinfo {author} {\bibfnamefont {C.}~\bibnamefont {Metzner}}, \bibinfo {author} {\bibfnamefont {K.}~\bibnamefont {Skodzek}}, \bibinfo {author} {\bibfnamefont {N.}~\bibnamefont {Lang}}, \bibinfo {author} {\bibfnamefont {I.}~\bibnamefont {Thievessen}}, \bibinfo {author} {\bibfnamefont {C.}~\bibnamefont {Mark}}, \bibinfo {author} {\bibfnamefont {S.}~\bibnamefont {M{\"u}nster}}, \bibinfo {author} {\bibfnamefont {K.~E.}\ \bibnamefont {Aifantis}},\ and\ \bibinfo {author} {\bibfnamefont {B.}~\bibnamefont {Fabry}},\ }\bibfield  {title} {\bibinfo {title} {Three-dimensional force microscopy of cells in biopolymer networks},\ }\href {https://doi.org/10.1038/nmeth.3685} {\bibfield  {journal} {\bibinfo  {journal} {Nature Methods}\ }\textbf {\bibinfo {volume} {13}},\ \bibinfo {pages} {171} (\bibinfo {year} {2016})}\BibitemShut {NoStop}%
\bibitem [{\citenamefont {Ban}\ \emph {et~al.}(2019)\citenamefont {Ban}, \citenamefont {Wang}, \citenamefont {Franklin}, \citenamefont {Liphardt}, \citenamefont {Janmey},\ and\ \citenamefont {Shenoy}}]{ban_strong_2019}%
  \BibitemOpen
  \bibfield  {author} {\bibinfo {author} {\bibfnamefont {E.}~\bibnamefont {Ban}}, \bibinfo {author} {\bibfnamefont {H.}~\bibnamefont {Wang}}, \bibinfo {author} {\bibfnamefont {J.~M.}\ \bibnamefont {Franklin}}, \bibinfo {author} {\bibfnamefont {J.~T.}\ \bibnamefont {Liphardt}}, \bibinfo {author} {\bibfnamefont {P.~A.}\ \bibnamefont {Janmey}},\ and\ \bibinfo {author} {\bibfnamefont {V.~B.}\ \bibnamefont {Shenoy}},\ }\bibfield  {title} {\bibinfo {title} {Strong triaxial coupling and anomalous {{Poisson}} effect in collagen networks},\ }\href {https://doi.org/10.1073/pnas.1815659116} {\bibfield  {journal} {\bibinfo  {journal} {Proceedings of the National Academy of Sciences}\ }\textbf {\bibinfo {volume} {116}},\ \bibinfo {pages} {6790} (\bibinfo {year} {2019})}\BibitemShut {NoStop}%
\bibitem [{\citenamefont {Poisson}(1827)}]{poisson_note_1827}%
  \BibitemOpen
  \bibfield  {author} {\bibinfo {author} {\bibfnamefont {S.~D.}\ \bibnamefont {Poisson}},\ }\bibfield  {title} {\bibinfo {title} {Note sur l'extension des fils et des plaques {\'e}lastiques},\ }\href@noop {} {\bibfield  {journal} {\bibinfo  {journal} {Annales de Chimie et de Physique}\ }\textbf {\bibinfo {volume} {36}},\ \bibinfo {pages} {384} (\bibinfo {year} {1827})}\BibitemShut {NoStop}%
\bibitem [{\citenamefont {Love}(1906)}]{love_treatise_1906}%
  \BibitemOpen
  \bibfield  {author} {\bibinfo {author} {\bibfnamefont {A.~E.~H.}\ \bibnamefont {Love}},\ }\href@noop {} {\emph {\bibinfo {title} {A {{Treatise}} on the {{Mathematical Theory}} of {{Elasticity}}}}},\ \bibinfo {edition} {2nd}\ ed.\ (\bibinfo  {publisher} {Cambridge University Press},\ \bibinfo {year} {1906})\BibitemShut {NoStop}%
\bibitem [{\citenamefont {Greaves}\ \emph {et~al.}(2011)\citenamefont {Greaves}, \citenamefont {Greer}, \citenamefont {Lakes},\ and\ \citenamefont {Rouxel}}]{greaves_poissons_2011}%
  \BibitemOpen
  \bibfield  {author} {\bibinfo {author} {\bibfnamefont {G.~N.}\ \bibnamefont {Greaves}}, \bibinfo {author} {\bibfnamefont {A.~L.}\ \bibnamefont {Greer}}, \bibinfo {author} {\bibfnamefont {R.~S.}\ \bibnamefont {Lakes}},\ and\ \bibinfo {author} {\bibfnamefont {T.}~\bibnamefont {Rouxel}},\ }\bibfield  {title} {\bibinfo {title} {Poisson's ratio and modern materials},\ }\href {https://doi.org/10.1038/nmat3134} {\bibfield  {journal} {\bibinfo  {journal} {Nature Materials}\ }\textbf {\bibinfo {volume} {10}},\ \bibinfo {pages} {823} (\bibinfo {year} {2011})}\BibitemShut {NoStop}%
\bibitem [{\citenamefont {Wolf}\ \emph {et~al.}(2013)\citenamefont {Wolf}, \citenamefont {{te Lindert}}, \citenamefont {Krause}, \citenamefont {Alexander}, \citenamefont {{te Riet}}, \citenamefont {Willis}, \citenamefont {Hoffman}, \citenamefont {Figdor}, \citenamefont {Weiss},\ and\ \citenamefont {Friedl}}]{wolf_physical_2013}%
  \BibitemOpen
  \bibfield  {author} {\bibinfo {author} {\bibfnamefont {K.}~\bibnamefont {Wolf}}, \bibinfo {author} {\bibfnamefont {M.}~\bibnamefont {{te Lindert}}}, \bibinfo {author} {\bibfnamefont {M.}~\bibnamefont {Krause}}, \bibinfo {author} {\bibfnamefont {S.}~\bibnamefont {Alexander}}, \bibinfo {author} {\bibfnamefont {J.}~\bibnamefont {{te Riet}}}, \bibinfo {author} {\bibfnamefont {A.~L.}\ \bibnamefont {Willis}}, \bibinfo {author} {\bibfnamefont {R.~M.}\ \bibnamefont {Hoffman}}, \bibinfo {author} {\bibfnamefont {C.~G.}\ \bibnamefont {Figdor}}, \bibinfo {author} {\bibfnamefont {S.~J.}\ \bibnamefont {Weiss}},\ and\ \bibinfo {author} {\bibfnamefont {P.}~\bibnamefont {Friedl}},\ }\bibfield  {title} {\bibinfo {title} {Physical limits of cell migration: {{Control}} by {{ECM}} space and nuclear deformation and tuning by proteolysis and traction force},\ }\href {https://doi.org/10.1083/jcb.201210152} {\bibfield  {journal} {\bibinfo  {journal} {The Journal of Cell Biology}\ }\textbf {\bibinfo {volume} {201}},\ \bibinfo
  {pages} {1069} (\bibinfo {year} {2013})}\BibitemShut {NoStop}%
\bibitem [{\citenamefont {Paul}\ \emph {et~al.}(2017)\citenamefont {Paul}, \citenamefont {Mistriotis},\ and\ \citenamefont {Konstantopoulos}}]{paul_cancer_2017}%
  \BibitemOpen
  \bibfield  {author} {\bibinfo {author} {\bibfnamefont {C.~D.}\ \bibnamefont {Paul}}, \bibinfo {author} {\bibfnamefont {P.}~\bibnamefont {Mistriotis}},\ and\ \bibinfo {author} {\bibfnamefont {K.}~\bibnamefont {Konstantopoulos}},\ }\bibfield  {title} {\bibinfo {title} {Cancer cell motility: Lessons from migration in confined spaces},\ }\href {https://doi.org/10.1038/nrc.2016.123} {\bibfield  {journal} {\bibinfo  {journal} {Nature Reviews Cancer}\ }\textbf {\bibinfo {volume} {17}},\ \bibinfo {pages} {131} (\bibinfo {year} {2017})}\BibitemShut {NoStop}%
\bibitem [{\citenamefont {Abhilash}\ \emph {et~al.}(2014)\citenamefont {Abhilash}, \citenamefont {Baker}, \citenamefont {Trappmann}, \citenamefont {Chen},\ and\ \citenamefont {Shenoy}}]{abhilash_remodeling_2014}%
  \BibitemOpen
  \bibfield  {author} {\bibinfo {author} {\bibfnamefont {A.~S.}\ \bibnamefont {Abhilash}}, \bibinfo {author} {\bibfnamefont {B.~M.}\ \bibnamefont {Baker}}, \bibinfo {author} {\bibfnamefont {B.}~\bibnamefont {Trappmann}}, \bibinfo {author} {\bibfnamefont {C.~S.}\ \bibnamefont {Chen}},\ and\ \bibinfo {author} {\bibfnamefont {V.~B.}\ \bibnamefont {Shenoy}},\ }\bibfield  {title} {\bibinfo {title} {Remodeling of fibrous extracellular matrices by contractile cells: {{Predictions}} from discrete fiber network simulations},\ }\href {https://doi.org/10.1016/j.bpj.2014.08.029} {\bibfield  {journal} {\bibinfo  {journal} {Biophysical Journal}\ }\textbf {\bibinfo {volume} {107}},\ \bibinfo {pages} {1829} (\bibinfo {year} {2014})}\BibitemShut {NoStop}%
\bibitem [{\citenamefont {Wang}\ \emph {et~al.}(2014)\citenamefont {Wang}, \citenamefont {Abhilash}, \citenamefont {Chen}, \citenamefont {Wells},\ and\ \citenamefont {Shenoy}}]{wang_long-range_2014}%
  \BibitemOpen
  \bibfield  {author} {\bibinfo {author} {\bibfnamefont {H.}~\bibnamefont {Wang}}, \bibinfo {author} {\bibfnamefont {A.~S.}\ \bibnamefont {Abhilash}}, \bibinfo {author} {\bibfnamefont {C.~S.}\ \bibnamefont {Chen}}, \bibinfo {author} {\bibfnamefont {R.~G.}\ \bibnamefont {Wells}},\ and\ \bibinfo {author} {\bibfnamefont {V.~B.}\ \bibnamefont {Shenoy}},\ }\bibfield  {title} {\bibinfo {title} {Long-range force transmission in fibrous matrices enabled by tension-driven alignment of fibers},\ }\href {https://doi.org/10.1016/j.bpj.2014.09.044} {\bibfield  {journal} {\bibinfo  {journal} {Biophysical Journal}\ }\textbf {\bibinfo {volume} {107}},\ \bibinfo {pages} {2592} (\bibinfo {year} {2014})}\BibitemShut {NoStop}%
\bibitem [{\citenamefont {Hall}\ \emph {et~al.}(2016)\citenamefont {Hall}, \citenamefont {Alisafaei}, \citenamefont {Ban}, \citenamefont {Feng}, \citenamefont {Hui}, \citenamefont {Shenoy},\ and\ \citenamefont {Wu}}]{hall_fibrous_2016}%
  \BibitemOpen
  \bibfield  {author} {\bibinfo {author} {\bibfnamefont {M.~S.}\ \bibnamefont {Hall}}, \bibinfo {author} {\bibfnamefont {F.}~\bibnamefont {Alisafaei}}, \bibinfo {author} {\bibfnamefont {E.}~\bibnamefont {Ban}}, \bibinfo {author} {\bibfnamefont {X.}~\bibnamefont {Feng}}, \bibinfo {author} {\bibfnamefont {C.-Y.}\ \bibnamefont {Hui}}, \bibinfo {author} {\bibfnamefont {V.~B.}\ \bibnamefont {Shenoy}},\ and\ \bibinfo {author} {\bibfnamefont {M.}~\bibnamefont {Wu}},\ }\bibfield  {title} {\bibinfo {title} {Fibrous nonlinear elasticity enables positive mechanical feedback between cells and {{ECMs}}},\ }\href {https://doi.org/10.1073/pnas.1613058113} {\bibfield  {journal} {\bibinfo  {journal} {Proceedings of the National Academy of Sciences}\ }\textbf {\bibinfo {volume} {113}},\ \bibinfo {pages} {14043} (\bibinfo {year} {2016})}\BibitemShut {NoStop}%
\bibitem [{\citenamefont {Notbohm}\ \emph {et~al.}(2015)\citenamefont {Notbohm}, \citenamefont {Lesman}, \citenamefont {Rosakis}, \citenamefont {Tirrell},\ and\ \citenamefont {Ravichandran}}]{notbohm_microbuckling_2015}%
  \BibitemOpen
  \bibfield  {author} {\bibinfo {author} {\bibfnamefont {J.}~\bibnamefont {Notbohm}}, \bibinfo {author} {\bibfnamefont {A.}~\bibnamefont {Lesman}}, \bibinfo {author} {\bibfnamefont {P.}~\bibnamefont {Rosakis}}, \bibinfo {author} {\bibfnamefont {D.~A.}\ \bibnamefont {Tirrell}},\ and\ \bibinfo {author} {\bibfnamefont {G.}~\bibnamefont {Ravichandran}},\ }\bibfield  {title} {\bibinfo {title} {Microbuckling of fibrin provides a mechanism for cell mechanosensing},\ }\href {https://doi.org/10.1098/rsif.2015.0320} {\bibfield  {journal} {\bibinfo  {journal} {Journal of the Royal Society, Interface}\ }\textbf {\bibinfo {volume} {12}},\ \bibinfo {pages} {20150320} (\bibinfo {year} {2015})}\BibitemShut {NoStop}%
\bibitem [{\citenamefont {SenGupta}\ \emph {et~al.}(2021)\citenamefont {SenGupta}, \citenamefont {Parent},\ and\ \citenamefont {Bear}}]{sengupta_principles_2021}%
  \BibitemOpen
  \bibfield  {author} {\bibinfo {author} {\bibfnamefont {S.}~\bibnamefont {SenGupta}}, \bibinfo {author} {\bibfnamefont {C.~A.}\ \bibnamefont {Parent}},\ and\ \bibinfo {author} {\bibfnamefont {J.~E.}\ \bibnamefont {Bear}},\ }\bibfield  {title} {\bibinfo {title} {The principles of directed cell migration},\ }\href {https://doi.org/10.1038/s41580-021-00366-6} {\bibfield  {journal} {\bibinfo  {journal} {Nature Reviews Molecular Cell Biology}\ }\textbf {\bibinfo {volume} {22}},\ \bibinfo {pages} {529} (\bibinfo {year} {2021})}\BibitemShut {NoStop}%
\bibitem [{\citenamefont {Ahmed}\ and\ \citenamefont {Saif}(2014)}]{ahmed_active_2014}%
  \BibitemOpen
  \bibfield  {author} {\bibinfo {author} {\bibfnamefont {W.~W.}\ \bibnamefont {Ahmed}}\ and\ \bibinfo {author} {\bibfnamefont {T.~A.}\ \bibnamefont {Saif}},\ }\bibfield  {title} {\bibinfo {title} {Active transport of vesicles in neurons is modulated by mechanical tension},\ }\href {https://doi.org/10.1038/srep04481} {\bibfield  {journal} {\bibinfo  {journal} {Scientific Reports}\ }\textbf {\bibinfo {volume} {4}},\ \bibinfo {pages} {4481} (\bibinfo {year} {2014})}\BibitemShut {NoStop}%
\bibitem [{\citenamefont {Svitkina}\ \emph {et~al.}(1997)\citenamefont {Svitkina}, \citenamefont {Verkhovsky}, \citenamefont {McQuade},\ and\ \citenamefont {Borisy}}]{svitkina_analysis_1997}%
  \BibitemOpen
  \bibfield  {author} {\bibinfo {author} {\bibfnamefont {T.~M.}\ \bibnamefont {Svitkina}}, \bibinfo {author} {\bibfnamefont {A.~B.}\ \bibnamefont {Verkhovsky}}, \bibinfo {author} {\bibfnamefont {K.~M.}\ \bibnamefont {McQuade}},\ and\ \bibinfo {author} {\bibfnamefont {G.~G.}\ \bibnamefont {Borisy}},\ }\bibfield  {title} {\bibinfo {title} {Analysis of the {{Actin}}--{{Myosin II System}} in {{Fish Epidermal Keratocytes}}: {{Mechanism}} of {{Cell Body Translocation}}},\ }\href {https://doi.org/10.1083/jcb.139.2.397} {\bibfield  {journal} {\bibinfo  {journal} {Journal of Cell Biology}\ }\textbf {\bibinfo {volume} {139}},\ \bibinfo {pages} {397} (\bibinfo {year} {1997})}\BibitemShut {NoStop}%
\bibitem [{\citenamefont {Keren}\ and\ \citenamefont {Theriot}(2008)}]{keren_biophysical_2008}%
  \BibitemOpen
  \bibfield  {author} {\bibinfo {author} {\bibfnamefont {K.}~\bibnamefont {Keren}}\ and\ \bibinfo {author} {\bibfnamefont {J.~A.}\ \bibnamefont {Theriot}},\ }\bibfield  {title} {\bibinfo {title} {Biophysical aspects of actin-based cell motility in fish epithelial keratocytes},\ }in\ \href@noop {} {\emph {\bibinfo {booktitle} {Cell {{Motility}}}}},\ \bibinfo {series and number} {Biological and {{Medical Physics}}, {{Biomedical Engineering}}},\ \bibinfo {editor} {edited by\ \bibinfo {editor} {\bibfnamefont {P.}~\bibnamefont {Lenz}}}\ (\bibinfo  {publisher} {Springer},\ \bibinfo {address} {New York, NY},\ \bibinfo {year} {2008})\ pp.\ \bibinfo {pages} {31--58}\BibitemShut {NoStop}%
\bibitem [{\citenamefont {Lee}\ \emph {et~al.}(2010)\citenamefont {Lee}, \citenamefont {Haase}, \citenamefont {Deguchi},\ and\ \citenamefont {Kaunas}}]{lee_cyclic_2010}%
  \BibitemOpen
  \bibfield  {author} {\bibinfo {author} {\bibfnamefont {C.-F.}\ \bibnamefont {Lee}}, \bibinfo {author} {\bibfnamefont {C.}~\bibnamefont {Haase}}, \bibinfo {author} {\bibfnamefont {S.}~\bibnamefont {Deguchi}},\ and\ \bibinfo {author} {\bibfnamefont {R.}~\bibnamefont {Kaunas}},\ }\bibfield  {title} {\bibinfo {title} {Cyclic stretch-induced stress fiber dynamics -- {{Dependence}} on strain rate, {{Rho-kinase}} and {{MLCK}}},\ }\href {https://doi.org/10.1016/j.bbrc.2010.09.046} {\bibfield  {journal} {\bibinfo  {journal} {Biochemical and Biophysical Research Communications}\ }\textbf {\bibinfo {volume} {401}},\ \bibinfo {pages} {344} (\bibinfo {year} {2010})}\BibitemShut {NoStop}%
\bibitem [{\citenamefont {Nagayama}\ \emph {et~al.}(2012)\citenamefont {Nagayama}, \citenamefont {Kimura}, \citenamefont {Makino},\ and\ \citenamefont {Matsumoto}}]{nagayama_strain_2012}%
  \BibitemOpen
  \bibfield  {author} {\bibinfo {author} {\bibfnamefont {K.}~\bibnamefont {Nagayama}}, \bibinfo {author} {\bibfnamefont {Y.}~\bibnamefont {Kimura}}, \bibinfo {author} {\bibfnamefont {N.}~\bibnamefont {Makino}},\ and\ \bibinfo {author} {\bibfnamefont {T.}~\bibnamefont {Matsumoto}},\ }\bibfield  {title} {\bibinfo {title} {Strain waveform dependence of stress fiber reorientation in cyclically stretched osteoblastic cells: Effects of viscoelastic compression of stress fibers},\ }\href {https://doi.org/10.1152/ajpcell.00155.2011} {\bibfield  {journal} {\bibinfo  {journal} {American Journal of Physiology-Cell Physiology}\ }\textbf {\bibinfo {volume} {302}},\ \bibinfo {pages} {C1469} (\bibinfo {year} {2012})}\BibitemShut {NoStop}%
\bibitem [{\citenamefont {Kaunas}(2015)}]{kaunas_dynamic_2015}%
  \BibitemOpen
  \bibfield  {author} {\bibinfo {author} {\bibfnamefont {R.}~\bibnamefont {Kaunas}},\ }\bibfield  {title} {\bibinfo {title} {Dynamic stress fiber reorganization on stretched matrices},\ }in\ \href@noop {} {\emph {\bibinfo {booktitle} {Cell and Matrix Mechanics}}},\ \bibinfo {editor} {edited by\ \bibinfo {editor} {\bibfnamefont {R.}~\bibnamefont {Kaunas}}\ and\ \bibinfo {editor} {\bibfnamefont {A.}~\bibnamefont {Zemel}}}\ (\bibinfo  {publisher} {CRC Press, Taylor \& Francis Group},\ \bibinfo {address} {Boca Raton},\ \bibinfo {year} {2015})\BibitemShut {NoStop}%
\bibitem [{\citenamefont {Kumar}\ \emph {et~al.}(2019)\citenamefont {Kumar}, \citenamefont {Shutova}, \citenamefont {Tanaka}, \citenamefont {Iwamoto}, \citenamefont {Calderwood}, \citenamefont {Svitkina},\ and\ \citenamefont {Schwartz}}]{kumar_filamin_2019}%
  \BibitemOpen
  \bibfield  {author} {\bibinfo {author} {\bibfnamefont {A.}~\bibnamefont {Kumar}}, \bibinfo {author} {\bibfnamefont {M.~S.}\ \bibnamefont {Shutova}}, \bibinfo {author} {\bibfnamefont {K.}~\bibnamefont {Tanaka}}, \bibinfo {author} {\bibfnamefont {D.~V.}\ \bibnamefont {Iwamoto}}, \bibinfo {author} {\bibfnamefont {D.~A.}\ \bibnamefont {Calderwood}}, \bibinfo {author} {\bibfnamefont {T.~M.}\ \bibnamefont {Svitkina}},\ and\ \bibinfo {author} {\bibfnamefont {M.~A.}\ \bibnamefont {Schwartz}},\ }\bibfield  {title} {\bibinfo {title} {Filamin {{A}} mediates isotropic distribution of applied force across the actin network},\ }\href {https://doi.org/10.1083/jcb.201901086} {\bibfield  {journal} {\bibinfo  {journal} {Journal of Cell Biology}\ }\textbf {\bibinfo {volume} {218}},\ \bibinfo {pages} {2481} (\bibinfo {year} {2019})}\BibitemShut {NoStop}%
\bibitem [{\citenamefont {Kabla}\ and\ \citenamefont {Mahadevan}(2007)}]{kabla_nonlinear_2007}%
  \BibitemOpen
  \bibfield  {author} {\bibinfo {author} {\bibfnamefont {A.}~\bibnamefont {Kabla}}\ and\ \bibinfo {author} {\bibfnamefont {L.}~\bibnamefont {Mahadevan}},\ }\bibfield  {title} {\bibinfo {title} {Nonlinear mechanics of soft fibrous networks},\ }\href {https://doi.org/10.1098/rsif.2006.0151} {\bibfield  {journal} {\bibinfo  {journal} {Journal of The Royal Society Interface}\ }\textbf {\bibinfo {volume} {4}},\ \bibinfo {pages} {99} (\bibinfo {year} {2007})}\BibitemShut {NoStop}%
\bibitem [{\citenamefont {Picu}\ \emph {et~al.}(2018)\citenamefont {Picu}, \citenamefont {Deogekar},\ and\ \citenamefont {Islam}}]{picu_poissons_2018}%
  \BibitemOpen
  \bibfield  {author} {\bibinfo {author} {\bibfnamefont {R.~C.}\ \bibnamefont {Picu}}, \bibinfo {author} {\bibfnamefont {S.}~\bibnamefont {Deogekar}},\ and\ \bibinfo {author} {\bibfnamefont {M.~R.}\ \bibnamefont {Islam}},\ }\bibfield  {title} {\bibinfo {title} {Poisson's {{Contraction}} and {{Fiber Kinematics}} in {{Tissue}}: {{Insight From Collagen Network Simulations}}},\ }\href {https://doi.org/10.1115/1.4038428} {\bibfield  {journal} {\bibinfo  {journal} {Journal of Biomechanical Engineering}\ }\textbf {\bibinfo {volume} {140}},\ \bibinfo {pages} {021002} (\bibinfo {year} {2018})}\BibitemShut {NoStop}%
\bibitem [{\citenamefont {Shivers}\ \emph {et~al.}(2020)\citenamefont {Shivers}, \citenamefont {Arzash},\ and\ \citenamefont {MacKintosh}}]{shivers_nonlinear_2020}%
  \BibitemOpen
  \bibfield  {author} {\bibinfo {author} {\bibfnamefont {J.~L.}\ \bibnamefont {Shivers}}, \bibinfo {author} {\bibfnamefont {S.}~\bibnamefont {Arzash}},\ and\ \bibinfo {author} {\bibfnamefont {F.~C.}\ \bibnamefont {MacKintosh}},\ }\bibfield  {title} {\bibinfo {title} {Nonlinear {{Poisson Effect Governed}} by a {{Mechanical Critical Transition}}},\ }\href {https://doi.org/10.1103/PhysRevLett.124.038002} {\bibfield  {journal} {\bibinfo  {journal} {Physical Review Letters}\ }\textbf {\bibinfo {volume} {124}},\ \bibinfo {pages} {038002} (\bibinfo {year} {2020})}\BibitemShut {NoStop}%
\bibitem [{\citenamefont {MacKintosh}\ \emph {et~al.}(1995)\citenamefont {MacKintosh}, \citenamefont {K{\"a}s},\ and\ \citenamefont {Janmey}}]{mackintosh_elasticity_1995}%
  \BibitemOpen
  \bibfield  {author} {\bibinfo {author} {\bibfnamefont {F.~C.}\ \bibnamefont {MacKintosh}}, \bibinfo {author} {\bibfnamefont {J.}~\bibnamefont {K{\"a}s}},\ and\ \bibinfo {author} {\bibfnamefont {P.~A.}\ \bibnamefont {Janmey}},\ }\bibfield  {title} {\bibinfo {title} {Elasticity of {{Semiflexible Biopolymer Networks}}},\ }\href {https://doi.org/10.1103/PhysRevLett.75.4425} {\bibfield  {journal} {\bibinfo  {journal} {Physical Review Letters}\ }\textbf {\bibinfo {volume} {75}},\ \bibinfo {pages} {4425} (\bibinfo {year} {1995})}\BibitemShut {NoStop}%
\bibitem [{\citenamefont {Odijk}(1995)}]{odijk_stiff_1995}%
  \BibitemOpen
  \bibfield  {author} {\bibinfo {author} {\bibfnamefont {T.}~\bibnamefont {Odijk}},\ }\bibfield  {title} {\bibinfo {title} {Stiff {{Chains}} and {{Filaments}} under {{Tension}}},\ }\href {https://doi.org/10.1021/ma00124a044} {\bibfield  {journal} {\bibinfo  {journal} {Macromolecules}\ }\textbf {\bibinfo {volume} {28}},\ \bibinfo {pages} {7016} (\bibinfo {year} {1995})}\BibitemShut {NoStop}%
\bibitem [{\citenamefont {Treloar}(1954)}]{treloar_photoelastic_1954}%
  \BibitemOpen
  \bibfield  {author} {\bibinfo {author} {\bibfnamefont {L.~R.~G.}\ \bibnamefont {Treloar}},\ }\bibfield  {title} {\bibinfo {title} {The photoelastic properties of short-chain molecular networks},\ }\href {https://doi.org/10.1039/TF9545000881} {\bibfield  {journal} {\bibinfo  {journal} {Transactions of the Faraday Society}\ }\textbf {\bibinfo {volume} {50}},\ \bibinfo {pages} {881} (\bibinfo {year} {1954})}\BibitemShut {NoStop}%
\bibitem [{\citenamefont {Treloar}(1975)}]{treloar_physics_1975}%
  \BibitemOpen
  \bibfield  {author} {\bibinfo {author} {\bibfnamefont {L.~R.~G.}\ \bibnamefont {Treloar}},\ }\href@noop {} {\emph {\bibinfo {title} {The {{Physics}} of {{Rubber Elasticity}}}}},\ \bibinfo {edition} {3rd}\ ed.\ (\bibinfo  {publisher} {Oxford University Press},\ \bibinfo {address} {Oxford},\ \bibinfo {year} {1975})\BibitemShut {NoStop}%
\bibitem [{\citenamefont {Treloar}\ \emph {et~al.}(1979)\citenamefont {Treloar}, \citenamefont {Riding},\ and\ \citenamefont {Gee}}]{treloar_non-gaussian_1979}%
  \BibitemOpen
  \bibfield  {author} {\bibinfo {author} {\bibfnamefont {L.~R.~G.}\ \bibnamefont {Treloar}}, \bibinfo {author} {\bibfnamefont {G.}~\bibnamefont {Riding}},\ and\ \bibinfo {author} {\bibfnamefont {G.}~\bibnamefont {Gee}},\ }\bibfield  {title} {\bibinfo {title} {A non-{{Gaussian}} theory for rubber in biaxial strain. {{I}}. {{Mechanical}} properties},\ }\href {https://doi.org/10.1098/rspa.1979.0163} {\bibfield  {journal} {\bibinfo  {journal} {Proceedings of the Royal Society of London. A}\ }\textbf {\bibinfo {volume} {369}},\ \bibinfo {pages} {261} (\bibinfo {year} {1979})}\BibitemShut {NoStop}%
\bibitem [{\citenamefont {Wu}\ and\ \citenamefont {{van der Giessen}}(1993)}]{wu_improved_1993}%
  \BibitemOpen
  \bibfield  {author} {\bibinfo {author} {\bibfnamefont {P.~D.}\ \bibnamefont {Wu}}\ and\ \bibinfo {author} {\bibfnamefont {E.}~\bibnamefont {{van der Giessen}}},\ }\bibfield  {title} {\bibinfo {title} {On improved network models for rubber elasticity and their applications to orientation hardening in glassy polymers},\ }\href {https://doi.org/10.1016/0022-5096(93)90043-F} {\bibfield  {journal} {\bibinfo  {journal} {Journal of the Mechanics and Physics of Solids}\ }\textbf {\bibinfo {volume} {41}},\ \bibinfo {pages} {427} (\bibinfo {year} {1993})}\BibitemShut {NoStop}%
\bibitem [{\citenamefont {Wu}\ and\ \citenamefont {{van der Giessen}}(1995)}]{wu_network_1995}%
  \BibitemOpen
  \bibfield  {author} {\bibinfo {author} {\bibfnamefont {P.~D.}\ \bibnamefont {Wu}}\ and\ \bibinfo {author} {\bibfnamefont {E.}~\bibnamefont {{van der Giessen}}},\ }\bibfield  {title} {\bibinfo {title} {On network descriptions of mechanical and optical properties of rubbers},\ }\href {https://doi.org/10.1080/01418619508236245} {\bibfield  {journal} {\bibinfo  {journal} {Philosophical Magazine A}\ }\textbf {\bibinfo {volume} {71}},\ \bibinfo {pages} {1191} (\bibinfo {year} {1995})}\BibitemShut {NoStop}%
\bibitem [{\citenamefont {Boyce}\ and\ \citenamefont {Arruda}(2000)}]{boyce_constitutive_2000}%
  \BibitemOpen
  \bibfield  {author} {\bibinfo {author} {\bibfnamefont {M.~C.}\ \bibnamefont {Boyce}}\ and\ \bibinfo {author} {\bibfnamefont {E.~M.}\ \bibnamefont {Arruda}},\ }\bibfield  {title} {\bibinfo {title} {Constitutive {{Models}} of {{Rubber Elasticity}}: {{A Review}}},\ }\href {https://doi.org/10.5254/1.3547602} {\bibfield  {journal} {\bibinfo  {journal} {Rubber Chemistry and Technology}\ }\textbf {\bibinfo {volume} {73}},\ \bibinfo {pages} {504} (\bibinfo {year} {2000})}\BibitemShut {NoStop}%
\bibitem [{\citenamefont {Palmer}\ and\ \citenamefont {Boyce}(2008)}]{palmer_constitutive_2008}%
  \BibitemOpen
  \bibfield  {author} {\bibinfo {author} {\bibfnamefont {J.~S.}\ \bibnamefont {Palmer}}\ and\ \bibinfo {author} {\bibfnamefont {M.~C.}\ \bibnamefont {Boyce}},\ }\bibfield  {title} {\bibinfo {title} {Constitutive modeling of the stress--strain behavior of {{F-actin}} filament networks},\ }\href {https://doi.org/10.1016/j.actbio.2007.12.007} {\bibfield  {journal} {\bibinfo  {journal} {Acta Biomaterialia}\ }\textbf {\bibinfo {volume} {4}},\ \bibinfo {pages} {597} (\bibinfo {year} {2008})}\BibitemShut {NoStop}%
\bibitem [{\citenamefont {Cioroianu}\ and\ \citenamefont {Storm}(2013)}]{cioroianu_normal_2013}%
  \BibitemOpen
  \bibfield  {author} {\bibinfo {author} {\bibfnamefont {A.~R.}\ \bibnamefont {Cioroianu}}\ and\ \bibinfo {author} {\bibfnamefont {C.}~\bibnamefont {Storm}},\ }\bibfield  {title} {\bibinfo {title} {Normal stresses in elastic networks},\ }\href {https://doi.org/10.1103/PhysRevE.88.052601} {\bibfield  {journal} {\bibinfo  {journal} {Physical Review E}\ }\textbf {\bibinfo {volume} {88}},\ \bibinfo {pages} {052601} (\bibinfo {year} {2013})}\BibitemShut {NoStop}%
\bibitem [{\citenamefont {Meng}\ and\ \citenamefont {Terentjev}(2017)}]{meng_theory_2017}%
  \BibitemOpen
  \bibfield  {author} {\bibinfo {author} {\bibfnamefont {F.}~\bibnamefont {Meng}}\ and\ \bibinfo {author} {\bibfnamefont {E.}~\bibnamefont {Terentjev}},\ }\bibfield  {title} {\bibinfo {title} {Theory of {{Semiflexible Filaments}} and {{Networks}}},\ }\href {https://doi.org/10.3390/polym9020052} {\bibfield  {journal} {\bibinfo  {journal} {Polymers}\ }\textbf {\bibinfo {volume} {9}},\ \bibinfo {pages} {52} (\bibinfo {year} {2017})}\BibitemShut {NoStop}%
\bibitem [{\citenamefont {Song}\ \emph {et~al.}(2022)\citenamefont {Song}, \citenamefont {Oberai},\ and\ \citenamefont {Janmey}}]{song_hyperelastic_2022}%
  \BibitemOpen
  \bibfield  {author} {\bibinfo {author} {\bibfnamefont {D.}~\bibnamefont {Song}}, \bibinfo {author} {\bibfnamefont {A.~A.}\ \bibnamefont {Oberai}},\ and\ \bibinfo {author} {\bibfnamefont {P.~A.}\ \bibnamefont {Janmey}},\ }\bibfield  {title} {\bibinfo {title} {Hyperelastic continuum models for isotropic athermal fibrous networks},\ }\href {https://doi.org/10.1098/rsfs.2022.0043} {\bibfield  {journal} {\bibinfo  {journal} {Interface Focus}\ }\textbf {\bibinfo {volume} {12}},\ \bibinfo {pages} {20220043} (\bibinfo {year} {2022})}\BibitemShut {NoStop}%
\bibitem [{\citenamefont {Rubinstein}\ and\ \citenamefont {Colby}(2006)}]{rubinstein_polymer_2006}%
  \BibitemOpen
  \bibfield  {author} {\bibinfo {author} {\bibfnamefont {M.}~\bibnamefont {Rubinstein}}\ and\ \bibinfo {author} {\bibfnamefont {R.~H.}\ \bibnamefont {Colby}},\ }\href@noop {} {\emph {\bibinfo {title} {Polymer {{Physics}}}}}\ (\bibinfo  {publisher} {Oxford University Press},\ \bibinfo {address} {Oxford},\ \bibinfo {year} {2006})\BibitemShut {NoStop}%
\bibitem [{\citenamefont {Vahabi}\ \emph {et~al.}(2018)\citenamefont {Vahabi}, \citenamefont {Vos}, \citenamefont {De~Cagny}, \citenamefont {Bonn}, \citenamefont {Koenderink},\ and\ \citenamefont {MacKintosh}}]{vahabi_normal_2018}%
  \BibitemOpen
  \bibfield  {author} {\bibinfo {author} {\bibfnamefont {M.}~\bibnamefont {Vahabi}}, \bibinfo {author} {\bibfnamefont {B.~E.}\ \bibnamefont {Vos}}, \bibinfo {author} {\bibfnamefont {H.~C.~G.}\ \bibnamefont {De~Cagny}}, \bibinfo {author} {\bibfnamefont {D.}~\bibnamefont {Bonn}}, \bibinfo {author} {\bibfnamefont {G.~H.}\ \bibnamefont {Koenderink}},\ and\ \bibinfo {author} {\bibfnamefont {F.~C.}\ \bibnamefont {MacKintosh}},\ }\bibfield  {title} {\bibinfo {title} {Normal stresses in semiflexible polymer hydrogels},\ }\href {https://doi.org/10.1103/PhysRevE.97.032418} {\bibfield  {journal} {\bibinfo  {journal} {Physical Review E}\ }\textbf {\bibinfo {volume} {97}},\ \bibinfo {pages} {032418} (\bibinfo {year} {2018})}\BibitemShut {NoStop}%
\bibitem [{Note1()}]{Note1}%
  \BibitemOpen
  \bibinfo {note} {For a homogeneous cylindrical elastic rod of radius $r$, the stretching modulus $\mu $ is related to the bending stiffness $\kappa $ as $\mu = 4\kappa /r^2$ \cite {storm_nonlinear_2005}.}\BibitemShut {Stop}%
\bibitem [{\citenamefont {Bustamante}\ \emph {et~al.}(1994)\citenamefont {Bustamante}, \citenamefont {Marko}, \citenamefont {Siggia},\ and\ \citenamefont {Smith}}]{bustamante_entropic_1994}%
  \BibitemOpen
  \bibfield  {author} {\bibinfo {author} {\bibfnamefont {C.}~\bibnamefont {Bustamante}}, \bibinfo {author} {\bibfnamefont {J.~F.}\ \bibnamefont {Marko}}, \bibinfo {author} {\bibfnamefont {E.~D.}\ \bibnamefont {Siggia}},\ and\ \bibinfo {author} {\bibfnamefont {S.}~\bibnamefont {Smith}},\ }\bibfield  {title} {\bibinfo {title} {Entropic {{Elasticity}} of {$\lambda$}-{{Phage DNA}}},\ }\href {https://doi.org/10.1126/science.8079175} {\bibfield  {journal} {\bibinfo  {journal} {Science}\ }\textbf {\bibinfo {volume} {265}},\ \bibinfo {pages} {1599} (\bibinfo {year} {1994})}\BibitemShut {NoStop}%
\bibitem [{\citenamefont {Fixman}\ and\ \citenamefont {Kovac}(1973)}]{fixman_polymer_1973}%
  \BibitemOpen
  \bibfield  {author} {\bibinfo {author} {\bibfnamefont {M.}~\bibnamefont {Fixman}}\ and\ \bibinfo {author} {\bibfnamefont {J.}~\bibnamefont {Kovac}},\ }\bibfield  {title} {\bibinfo {title} {Polymer conformational statistics. {{III}}. {{Modified Gaussian}} models of stiff chains},\ }\href {https://doi.org/10.1063/1.1679396} {\bibfield  {journal} {\bibinfo  {journal} {The Journal of Chemical Physics}\ }\textbf {\bibinfo {volume} {58}},\ \bibinfo {pages} {1564} (\bibinfo {year} {1973})}\BibitemShut {NoStop}%
\bibitem [{\citenamefont {James}\ and\ \citenamefont {Guth}(1943)}]{james_theory_1943}%
  \BibitemOpen
  \bibfield  {author} {\bibinfo {author} {\bibfnamefont {H.~M.}\ \bibnamefont {James}}\ and\ \bibinfo {author} {\bibfnamefont {E.}~\bibnamefont {Guth}},\ }\bibfield  {title} {\bibinfo {title} {Theory of the {{Elastic Properties}} of {{Rubber}}},\ }\href {https://doi.org/10.1063/1.1723785} {\bibfield  {journal} {\bibinfo  {journal} {The Journal of Chemical Physics}\ }\textbf {\bibinfo {volume} {11}},\ \bibinfo {pages} {455} (\bibinfo {year} {1943})}\BibitemShut {NoStop}%
\bibitem [{\citenamefont {Wu}\ and\ \citenamefont {{van der Giessen}}(1992)}]{wu_improved_1992}%
  \BibitemOpen
  \bibfield  {author} {\bibinfo {author} {\bibfnamefont {P.~D.}\ \bibnamefont {Wu}}\ and\ \bibinfo {author} {\bibfnamefont {E.}~\bibnamefont {{van der Giessen}}},\ }\bibfield  {title} {\bibinfo {title} {On improved 3-{{D}} non-{{Gaussian}} network models for rubber elasticity},\ }\href {https://doi.org/10.1016/0093-6413(92)90021-2} {\bibfield  {journal} {\bibinfo  {journal} {Mechanics Research Communications}\ }\textbf {\bibinfo {volume} {19}},\ \bibinfo {pages} {427} (\bibinfo {year} {1992})}\BibitemShut {NoStop}%
\bibitem [{\citenamefont {Kroy}\ and\ \citenamefont {Frey}(1996)}]{kroy_force-extension_1996}%
  \BibitemOpen
  \bibfield  {author} {\bibinfo {author} {\bibfnamefont {K.}~\bibnamefont {Kroy}}\ and\ \bibinfo {author} {\bibfnamefont {E.}~\bibnamefont {Frey}},\ }\bibfield  {title} {\bibinfo {title} {Force-{{Extension Relation}} and {{Plateau Modulus}} for {{Wormlike Chains}}},\ }\href {https://doi.org/10.1103/PhysRevLett.77.306} {\bibfield  {journal} {\bibinfo  {journal} {Physical Review Letters}\ }\textbf {\bibinfo {volume} {77}},\ \bibinfo {pages} {306} (\bibinfo {year} {1996})}\BibitemShut {NoStop}%
\bibitem [{\citenamefont {Broedersz}\ \emph {et~al.}(2009)\citenamefont {Broedersz}, \citenamefont {Storm},\ and\ \citenamefont {MacKintosh}}]{broedersz_effective-medium_2009}%
  \BibitemOpen
  \bibfield  {author} {\bibinfo {author} {\bibfnamefont {C.~P.}\ \bibnamefont {Broedersz}}, \bibinfo {author} {\bibfnamefont {C.}~\bibnamefont {Storm}},\ and\ \bibinfo {author} {\bibfnamefont {F.~C.}\ \bibnamefont {MacKintosh}},\ }\bibfield  {title} {\bibinfo {title} {Effective-medium approach for stiff polymer networks with flexible cross-links},\ }\href {https://doi.org/10.1103/physreve.79.061914} {\bibfield  {journal} {\bibinfo  {journal} {Physical Review E}\ }\textbf {\bibinfo {volume} {79}},\ \bibinfo {pages} {061914} (\bibinfo {year} {2009})}\BibitemShut {NoStop}%
\bibitem [{\citenamefont {Lin}\ \emph {et~al.}(2010)\citenamefont {Lin}, \citenamefont {Yao}, \citenamefont {Broedersz}, \citenamefont {Herrmann}, \citenamefont {MacKintosh},\ and\ \citenamefont {Weitz}}]{lin_origins_2010}%
  \BibitemOpen
  \bibfield  {author} {\bibinfo {author} {\bibfnamefont {Y.-C.}\ \bibnamefont {Lin}}, \bibinfo {author} {\bibfnamefont {N.~Y.}\ \bibnamefont {Yao}}, \bibinfo {author} {\bibfnamefont {C.~P.}\ \bibnamefont {Broedersz}}, \bibinfo {author} {\bibfnamefont {H.}~\bibnamefont {Herrmann}}, \bibinfo {author} {\bibfnamefont {F.~C.}\ \bibnamefont {MacKintosh}},\ and\ \bibinfo {author} {\bibfnamefont {D.~A.}\ \bibnamefont {Weitz}},\ }\bibfield  {title} {\bibinfo {title} {Origins of {{Elasticity}} in {{Intermediate Filament Networks}}},\ }\href {https://doi.org/10.1103/PhysRevLett.104.058101} {\bibfield  {journal} {\bibinfo  {journal} {Physical Review Letters}\ }\textbf {\bibinfo {volume} {104}},\ \bibinfo {pages} {058101} (\bibinfo {year} {2010})}\BibitemShut {NoStop}%
\bibitem [{\citenamefont {Yao}\ \emph {et~al.}(2010)\citenamefont {Yao}, \citenamefont {Broedersz}, \citenamefont {Lin}, \citenamefont {Kasza}, \citenamefont {MacKintosh},\ and\ \citenamefont {Weitz}}]{yao_elasticity_2010}%
  \BibitemOpen
  \bibfield  {author} {\bibinfo {author} {\bibfnamefont {N.~Y.}\ \bibnamefont {Yao}}, \bibinfo {author} {\bibfnamefont {C.~P.}\ \bibnamefont {Broedersz}}, \bibinfo {author} {\bibfnamefont {Y.-C.}\ \bibnamefont {Lin}}, \bibinfo {author} {\bibfnamefont {K.~E.}\ \bibnamefont {Kasza}}, \bibinfo {author} {\bibfnamefont {F.~C.}\ \bibnamefont {MacKintosh}},\ and\ \bibinfo {author} {\bibfnamefont {D.~A.}\ \bibnamefont {Weitz}},\ }\bibfield  {title} {\bibinfo {title} {Elasticity in {{Ionically Cross-Linked Neurofilament Networks}}},\ }\href {https://doi.org/10.1016/j.bpj.2010.01.062} {\bibfield  {journal} {\bibinfo  {journal} {Biophysical Journal}\ }\textbf {\bibinfo {volume} {98}},\ \bibinfo {pages} {2147} (\bibinfo {year} {2010})}\BibitemShut {NoStop}%
\bibitem [{\citenamefont {Van~Oosterwyck}\ \emph {et~al.}(2013)\citenamefont {Van~Oosterwyck}, \citenamefont {Rodr{\'i}guez}, \citenamefont {Doblar{\'e}},\ and\ \citenamefont {Garc{\'i}a~Aznar}}]{van_oosterwyck_affine_2013}%
  \BibitemOpen
  \bibfield  {author} {\bibinfo {author} {\bibfnamefont {H.}~\bibnamefont {Van~Oosterwyck}}, \bibinfo {author} {\bibfnamefont {J.~F.}\ \bibnamefont {Rodr{\'i}guez}}, \bibinfo {author} {\bibfnamefont {M.}~\bibnamefont {Doblar{\'e}}},\ and\ \bibinfo {author} {\bibfnamefont {J.~M.}\ \bibnamefont {Garc{\'i}a~Aznar}},\ }\bibfield  {title} {\bibinfo {title} {An affine micro-sphere-based constitutive model, accounting for junctional sliding, can capture {{F-actin}} network mechanics},\ }\href {https://doi.org/10.1080/10255842.2011.648626} {\bibfield  {journal} {\bibinfo  {journal} {Computer Methods in Biomechanics and Biomedical Engineering}\ }\textbf {\bibinfo {volume} {16}},\ \bibinfo {pages} {1002} (\bibinfo {year} {2013})}\BibitemShut {NoStop}%
\bibitem [{\citenamefont {Holzapfel}\ \emph {et~al.}(2014)\citenamefont {Holzapfel}, \citenamefont {Unterberger},\ and\ \citenamefont {Ogden}}]{holzapfel_affine_2014}%
  \BibitemOpen
  \bibfield  {author} {\bibinfo {author} {\bibfnamefont {G.~A.}\ \bibnamefont {Holzapfel}}, \bibinfo {author} {\bibfnamefont {M.~J.}\ \bibnamefont {Unterberger}},\ and\ \bibinfo {author} {\bibfnamefont {R.~W.}\ \bibnamefont {Ogden}},\ }\bibfield  {title} {\bibinfo {title} {An affine continuum mechanical model for cross-linked {{F-actin}} networks with compliant linker proteins},\ }\href {https://doi.org/10.1016/j.jmbbm.2014.05.014} {\bibfield  {journal} {\bibinfo  {journal} {Journal of the Mechanical Behavior of Biomedical Materials}\ }\textbf {\bibinfo {volume} {38}},\ \bibinfo {pages} {78} (\bibinfo {year} {2014})}\BibitemShut {NoStop}%
\bibitem [{\citenamefont {Unterberger}\ and\ \citenamefont {Holzapfel}(2014)}]{unterberger_advances_2014}%
  \BibitemOpen
  \bibfield  {author} {\bibinfo {author} {\bibfnamefont {M.~J.}\ \bibnamefont {Unterberger}}\ and\ \bibinfo {author} {\bibfnamefont {G.~A.}\ \bibnamefont {Holzapfel}},\ }\bibfield  {title} {\bibinfo {title} {Advances in the mechanical modeling of filamentous actin and its cross-linked networks on multiple scales},\ }\href {https://doi.org/10.1007/s10237-014-0578-4} {\bibfield  {journal} {\bibinfo  {journal} {Biomechanics and Modeling in Mechanobiology}\ }\textbf {\bibinfo {volume} {13}},\ \bibinfo {pages} {1155} (\bibinfo {year} {2014})}\BibitemShut {NoStop}%
\bibitem [{\citenamefont {Larson}(1999)}]{larson_structure_1999}%
  \BibitemOpen
  \bibfield  {author} {\bibinfo {author} {\bibfnamefont {R.~G.}\ \bibnamefont {Larson}},\ }\href@noop {} {\emph {\bibinfo {title} {The {{Structure}} and {{Rheology}} of {{Complex Fluids}}}}}\ (\bibinfo  {publisher} {Oxford University Press},\ \bibinfo {year} {1999})\BibitemShut {NoStop}%
\bibitem [{\citenamefont {Morse}(1998)}]{morse_viscoelasticity_1998-2}%
  \BibitemOpen
  \bibfield  {author} {\bibinfo {author} {\bibfnamefont {D.~C.}\ \bibnamefont {Morse}},\ }\bibfield  {title} {\bibinfo {title} {Viscoelasticity of {{Concentrated Isotropic Solutions}} of {{Semiflexible Polymers}}. 1. {{Model}} and {{Stress Tensor}}},\ }\href {https://doi.org/10.1021/ma9803032} {\bibfield  {journal} {\bibinfo  {journal} {Macromolecules}\ }\textbf {\bibinfo {volume} {31}},\ \bibinfo {pages} {7030} (\bibinfo {year} {1998})}\BibitemShut {NoStop}%
\bibitem [{\citenamefont {Gittes}\ and\ \citenamefont {MacKintosh}(1998)}]{gittes_dynamic_1998}%
  \BibitemOpen
  \bibfield  {author} {\bibinfo {author} {\bibfnamefont {F.}~\bibnamefont {Gittes}}\ and\ \bibinfo {author} {\bibfnamefont {F.~C.}\ \bibnamefont {MacKintosh}},\ }\bibfield  {title} {\bibinfo {title} {Dynamic shear modulus of a semiflexible polymer network},\ }\href {https://doi.org/10.1103/PhysRevE.58.R1241} {\bibfield  {journal} {\bibinfo  {journal} {Physical Review E}\ }\textbf {\bibinfo {volume} {58}},\ \bibinfo {pages} {R1241} (\bibinfo {year} {1998})}\BibitemShut {NoStop}%
\bibitem [{\citenamefont {{de Gennes}}\ and\ \citenamefont {Prost}(2013)}]{de_gennes_physics_2013}%
  \BibitemOpen
  \bibfield  {author} {\bibinfo {author} {\bibfnamefont {P.-G.}\ \bibnamefont {{de Gennes}}}\ and\ \bibinfo {author} {\bibfnamefont {J.}~\bibnamefont {Prost}},\ }\href@noop {} {\emph {\bibinfo {title} {The {{Physics}} of {{Liquid Crystals}}}}},\ \bibinfo {number} {83}\ (\bibinfo  {publisher} {Clarendon Press},\ \bibinfo {address} {Oxford},\ \bibinfo {year} {2013})\BibitemShut {NoStop}%
\bibitem [{\citenamefont {Feng}\ \emph {et~al.}(2015)\citenamefont {Feng}, \citenamefont {Levine}, \citenamefont {Mao},\ and\ \citenamefont {Sander}}]{feng_alignment_2015}%
  \BibitemOpen
  \bibfield  {author} {\bibinfo {author} {\bibfnamefont {J.}~\bibnamefont {Feng}}, \bibinfo {author} {\bibfnamefont {H.}~\bibnamefont {Levine}}, \bibinfo {author} {\bibfnamefont {X.}~\bibnamefont {Mao}},\ and\ \bibinfo {author} {\bibfnamefont {L.~M.}\ \bibnamefont {Sander}},\ }\bibfield  {title} {\bibinfo {title} {Alignment and nonlinear elasticity in biopolymer gels},\ }\href {https://doi.org/10.1103/PhysRevE.91.042710} {\bibfield  {journal} {\bibinfo  {journal} {Physical Review E}\ }\textbf {\bibinfo {volume} {91}},\ \bibinfo {pages} {042710} (\bibinfo {year} {2015})}\BibitemShut {NoStop}%
\bibitem [{\citenamefont {Doi}(2004)}]{doi_introduction_2004}%
  \BibitemOpen
  \bibfield  {author} {\bibinfo {author} {\bibfnamefont {M.}~\bibnamefont {Doi}},\ }\href@noop {} {\emph {\bibinfo {title} {Introduction to Polymer Physics}}}\ (\bibinfo  {publisher} {Clarendon Press},\ \bibinfo {address} {Oxford},\ \bibinfo {year} {2004})\BibitemShut {NoStop}%
\bibitem [{\citenamefont {Macosko}(1994)}]{macosko_rheology_1994}%
  \BibitemOpen
  \bibfield  {author} {\bibinfo {author} {\bibfnamefont {C.~W.}\ \bibnamefont {Macosko}},\ }\href@noop {} {\emph {\bibinfo {title} {Rheology: Principles, Measurements, and Applications}}},\ Advances in Interfacial Engineering Series\ (\bibinfo  {publisher} {VCH},\ \bibinfo {address} {New York, NY},\ \bibinfo {year} {1994})\BibitemShut {NoStop}%
\bibitem [{\citenamefont {Das}\ and\ \citenamefont {MacKintosh}(2010)}]{das_poissons_2010}%
  \BibitemOpen
  \bibfield  {author} {\bibinfo {author} {\bibfnamefont {M.}~\bibnamefont {Das}}\ and\ \bibinfo {author} {\bibfnamefont {F.~C.}\ \bibnamefont {MacKintosh}},\ }\bibfield  {title} {\bibinfo {title} {Poisson's {{Ratio}} in {{Composite Elastic Media}} with {{Rigid Rods}}},\ }\href {https://doi.org/10.1103/PhysRevLett.105.138102} {\bibfield  {journal} {\bibinfo  {journal} {Physical Review Letters}\ }\textbf {\bibinfo {volume} {105}},\ \bibinfo {pages} {138102} (\bibinfo {year} {2010})}\BibitemShut {NoStop}%
\bibitem [{\citenamefont {Odijk}(1998)}]{odijk_microfibrillar_1998}%
  \BibitemOpen
  \bibfield  {author} {\bibinfo {author} {\bibfnamefont {T.}~\bibnamefont {Odijk}},\ }\bibfield  {title} {\bibinfo {title} {Microfibrillar buckling within fibers under compression},\ }\href {https://doi.org/10.1063/1.476107} {\bibfield  {journal} {\bibinfo  {journal} {The Journal of Chemical Physics}\ }\textbf {\bibinfo {volume} {108}},\ \bibinfo {pages} {6923} (\bibinfo {year} {1998})}\BibitemShut {NoStop}%
\bibitem [{\citenamefont {Emanuel}\ \emph {et~al.}(2007)\citenamefont {Emanuel}, \citenamefont {Mohrbach}, \citenamefont {Sayar}, \citenamefont {Schiessel},\ and\ \citenamefont {Kuli{\'c}}}]{emanuel_buckling_2007}%
  \BibitemOpen
  \bibfield  {author} {\bibinfo {author} {\bibfnamefont {M.}~\bibnamefont {Emanuel}}, \bibinfo {author} {\bibfnamefont {H.}~\bibnamefont {Mohrbach}}, \bibinfo {author} {\bibfnamefont {M.}~\bibnamefont {Sayar}}, \bibinfo {author} {\bibfnamefont {H.}~\bibnamefont {Schiessel}},\ and\ \bibinfo {author} {\bibfnamefont {I.~M.}\ \bibnamefont {Kuli{\'c}}},\ }\bibfield  {title} {\bibinfo {title} {Buckling of stiff polymers: {{Influence}} of thermal fluctuations},\ }\href {https://doi.org/10.1103/PhysRevE.76.061907} {\bibfield  {journal} {\bibinfo  {journal} {Physical Review E}\ }\textbf {\bibinfo {volume} {76}},\ \bibinfo {pages} {061907} (\bibinfo {year} {2007})}\BibitemShut {NoStop}%
\bibitem [{\citenamefont {Baczynski}\ \emph {et~al.}(2008)\citenamefont {Baczynski}, \citenamefont {Lipowsky},\ and\ \citenamefont {Kierfeld}}]{baczynski_stretching_2008}%
  \BibitemOpen
  \bibfield  {author} {\bibinfo {author} {\bibfnamefont {K.}~\bibnamefont {Baczynski}}, \bibinfo {author} {\bibfnamefont {R.}~\bibnamefont {Lipowsky}},\ and\ \bibinfo {author} {\bibfnamefont {J.}~\bibnamefont {Kierfeld}},\ }\bibfield  {title} {\bibinfo {title} {Stretching of buckled filaments by thermal fluctuations},\ }\href {https://doi.org/10.1103/physreve.76.061914} {\bibfield  {journal} {\bibinfo  {journal} {Physical Review E}\ }\textbf {\bibinfo {volume} {76}},\ \bibinfo {pages} {1} (\bibinfo {year} {2008})}\BibitemShut {NoStop}%
\bibitem [{\citenamefont {Blundell}\ and\ \citenamefont {Terentjev}(2009)}]{blundell_buckling_2009}%
  \BibitemOpen
  \bibfield  {author} {\bibinfo {author} {\bibfnamefont {J.~R.}\ \bibnamefont {Blundell}}\ and\ \bibinfo {author} {\bibfnamefont {E.~M.}\ \bibnamefont {Terentjev}},\ }\bibfield  {title} {\bibinfo {title} {Buckling of semiflexible filaments under compression},\ }\href {https://doi.org/10.1039/B903583D} {\bibfield  {journal} {\bibinfo  {journal} {Soft Matter}\ }\textbf {\bibinfo {volume} {5}},\ \bibinfo {pages} {4015} (\bibinfo {year} {2009})}\BibitemShut {NoStop}%
\bibitem [{\citenamefont {Kurzthaler}(2018)}]{kurzthaler_elastic_2018}%
  \BibitemOpen
  \bibfield  {author} {\bibinfo {author} {\bibfnamefont {C.}~\bibnamefont {Kurzthaler}},\ }\bibfield  {title} {\bibinfo {title} {Elastic behavior of a semiflexible polymer in {{3D}} subject to compression and stretching forces},\ }\href {https://doi.org/10.1039/C8SM01403E} {\bibfield  {journal} {\bibinfo  {journal} {Soft Matter}\ }\textbf {\bibinfo {volume} {14}},\ \bibinfo {pages} {7634} (\bibinfo {year} {2018})}\BibitemShut {NoStop}%
\bibitem [{\citenamefont {Unterberger}\ \emph {et~al.}(2013)\citenamefont {Unterberger}, \citenamefont {Schmoller}, \citenamefont {Bausch},\ and\ \citenamefont {Holzapfel}}]{unterberger_new_2013}%
  \BibitemOpen
  \bibfield  {author} {\bibinfo {author} {\bibfnamefont {M.~J.}\ \bibnamefont {Unterberger}}, \bibinfo {author} {\bibfnamefont {K.~M.}\ \bibnamefont {Schmoller}}, \bibinfo {author} {\bibfnamefont {A.~R.}\ \bibnamefont {Bausch}},\ and\ \bibinfo {author} {\bibfnamefont {G.~A.}\ \bibnamefont {Holzapfel}},\ }\bibfield  {title} {\bibinfo {title} {A new approach to model cross-linked actin networks: {{Multi-scale}} continuum formulation and computational analysis},\ }\href {https://doi.org/10.1016/j.jmbbm.2012.11.019} {\bibfield  {journal} {\bibinfo  {journal} {Journal of the Mechanical Behavior of Biomedical Materials}\ }\textbf {\bibinfo {volume} {22}},\ \bibinfo {pages} {95} (\bibinfo {year} {2013})}\BibitemShut {NoStop}%
\bibitem [{\citenamefont {Holzapfel}\ and\ \citenamefont {Ogden}(2013)}]{holzapfel_elasticity_2013}%
  \BibitemOpen
  \bibfield  {author} {\bibinfo {author} {\bibfnamefont {G.~A.}\ \bibnamefont {Holzapfel}}\ and\ \bibinfo {author} {\bibfnamefont {R.~W.}\ \bibnamefont {Ogden}},\ }\bibfield  {title} {\bibinfo {title} {Elasticity of biopolymer filaments},\ }\href {https://doi.org/10.1016/j.actbio.2013.03.001} {\bibfield  {journal} {\bibinfo  {journal} {Acta Biomaterialia}\ }\textbf {\bibinfo {volume} {9}},\ \bibinfo {pages} {7320} (\bibinfo {year} {2013})}\BibitemShut {NoStop}%
\bibitem [{\citenamefont {Li}\ \emph {et~al.}(2022)\citenamefont {Li}, \citenamefont {Li}, \citenamefont {Prince}, \citenamefont {Weitz}, \citenamefont {Panyukov}, \citenamefont {Ramachandran}, \citenamefont {Rubinstein},\ and\ \citenamefont {Kumacheva}}]{li_fibrous_2022}%
  \BibitemOpen
  \bibfield  {author} {\bibinfo {author} {\bibfnamefont {Y.}~\bibnamefont {Li}}, \bibinfo {author} {\bibfnamefont {Y.}~\bibnamefont {Li}}, \bibinfo {author} {\bibfnamefont {E.}~\bibnamefont {Prince}}, \bibinfo {author} {\bibfnamefont {J.~I.}\ \bibnamefont {Weitz}}, \bibinfo {author} {\bibfnamefont {S.}~\bibnamefont {Panyukov}}, \bibinfo {author} {\bibfnamefont {A.}~\bibnamefont {Ramachandran}}, \bibinfo {author} {\bibfnamefont {M.}~\bibnamefont {Rubinstein}},\ and\ \bibinfo {author} {\bibfnamefont {E.}~\bibnamefont {Kumacheva}},\ }\bibfield  {title} {\bibinfo {title} {Fibrous hydrogels under biaxial confinement},\ }\href {https://doi.org/10.1038/s41467-022-30980-7} {\bibfield  {journal} {\bibinfo  {journal} {Nature Communications}\ }\textbf {\bibinfo {volume} {13}},\ \bibinfo {pages} {3264} (\bibinfo {year} {2022})}\BibitemShut {NoStop}%
\bibitem [{\citenamefont {Janmey}\ \emph {et~al.}(2007)\citenamefont {Janmey}, \citenamefont {McCormick}, \citenamefont {Rammensee}, \citenamefont {Leight}, \citenamefont {Georges},\ and\ \citenamefont {MacKintosh}}]{janmey_negative_2007}%
  \BibitemOpen
  \bibfield  {author} {\bibinfo {author} {\bibfnamefont {P.~A.}\ \bibnamefont {Janmey}}, \bibinfo {author} {\bibfnamefont {M.~E.}\ \bibnamefont {McCormick}}, \bibinfo {author} {\bibfnamefont {S.}~\bibnamefont {Rammensee}}, \bibinfo {author} {\bibfnamefont {J.~L.}\ \bibnamefont {Leight}}, \bibinfo {author} {\bibfnamefont {P.~C.}\ \bibnamefont {Georges}},\ and\ \bibinfo {author} {\bibfnamefont {F.~C.}\ \bibnamefont {MacKintosh}},\ }\bibfield  {title} {\bibinfo {title} {Negative normal stress in semiflexible biopolymer gels},\ }\href {https://doi.org/10.1038/nmat1810} {\bibfield  {journal} {\bibinfo  {journal} {Nature Materials}\ }\textbf {\bibinfo {volume} {6}},\ \bibinfo {pages} {48} (\bibinfo {year} {2007})}\BibitemShut {NoStop}%
\bibitem [{\citenamefont {De~Cagny}\ \emph {et~al.}(2016)\citenamefont {De~Cagny}, \citenamefont {Vos}, \citenamefont {Vahabi}, \citenamefont {Kurniawan}, \citenamefont {Doi}, \citenamefont {Koenderink}, \citenamefont {MacKintosh},\ and\ \citenamefont {Bonn}}]{de_cagny_porosity_2016}%
  \BibitemOpen
  \bibfield  {author} {\bibinfo {author} {\bibfnamefont {H.~C.~G.}\ \bibnamefont {De~Cagny}}, \bibinfo {author} {\bibfnamefont {B.~E.}\ \bibnamefont {Vos}}, \bibinfo {author} {\bibfnamefont {M.}~\bibnamefont {Vahabi}}, \bibinfo {author} {\bibfnamefont {N.~A.}\ \bibnamefont {Kurniawan}}, \bibinfo {author} {\bibfnamefont {M.}~\bibnamefont {Doi}}, \bibinfo {author} {\bibfnamefont {G.~H.}\ \bibnamefont {Koenderink}}, \bibinfo {author} {\bibfnamefont {F.~C.}\ \bibnamefont {MacKintosh}},\ and\ \bibinfo {author} {\bibfnamefont {D.}~\bibnamefont {Bonn}},\ }\bibfield  {title} {\bibinfo {title} {Porosity {{Governs Normal Stresses}} in {{Polymer Gels}}},\ }\href {https://doi.org/10.1103/PhysRevLett.117.217802} {\bibfield  {journal} {\bibinfo  {journal} {Physical Review Letters}\ }\textbf {\bibinfo {volume} {117}},\ \bibinfo {pages} {217802} (\bibinfo {year} {2016})}\BibitemShut {NoStop}%
\bibitem [{\citenamefont {Doi}(2013)}]{doi_soft_2013}%
  \BibitemOpen
  \bibfield  {author} {\bibinfo {author} {\bibfnamefont {M.}~\bibnamefont {Doi}},\ }\href@noop {} {\emph {\bibinfo {title} {Soft {{Matter Physics}}}}}\ (\bibinfo  {publisher} {Oxford University Press},\ \bibinfo {year} {2013})\BibitemShut {NoStop}%
\bibitem [{\citenamefont {Yamamoto}\ \emph {et~al.}(2017)\citenamefont {Yamamoto}, \citenamefont {Masubuchi},\ and\ \citenamefont {Doi}}]{yamamoto_large_2017}%
  \BibitemOpen
  \bibfield  {author} {\bibinfo {author} {\bibfnamefont {T.}~\bibnamefont {Yamamoto}}, \bibinfo {author} {\bibfnamefont {Y.}~\bibnamefont {Masubuchi}},\ and\ \bibinfo {author} {\bibfnamefont {M.}~\bibnamefont {Doi}},\ }\bibfield  {title} {\bibinfo {title} {Large {{Network Swelling}} and {{Solvent Redistribution Are Necessary}} for {{Polymer Gels}} to {{Show Negative Normal Stress}}},\ }\href {https://doi.org/10.1021/acsmacrolett.7b00153} {\bibfield  {journal} {\bibinfo  {journal} {ACS Macro Letters}\ }\textbf {\bibinfo {volume} {6}},\ \bibinfo {pages} {512} (\bibinfo {year} {2017})}\BibitemShut {NoStop}%
\bibitem [{\citenamefont {Punter}\ \emph {et~al.}(2020)\citenamefont {Punter}, \citenamefont {Vos}, \citenamefont {Mulder},\ and\ \citenamefont {Koenderink}}]{punter_poroelasticity_2020}%
  \BibitemOpen
  \bibfield  {author} {\bibinfo {author} {\bibfnamefont {M.~T. J. J.~M.}\ \bibnamefont {Punter}}, \bibinfo {author} {\bibfnamefont {B.~E.}\ \bibnamefont {Vos}}, \bibinfo {author} {\bibfnamefont {B.~M.}\ \bibnamefont {Mulder}},\ and\ \bibinfo {author} {\bibfnamefont {G.~H.}\ \bibnamefont {Koenderink}},\ }\bibfield  {title} {\bibinfo {title} {Poroelasticity of (bio)polymer networks during compression: Theory and experiment},\ }\href {https://doi.org/10.1039/C9SM01973A} {\bibfield  {journal} {\bibinfo  {journal} {Soft Matter}\ }\textbf {\bibinfo {volume} {16}},\ \bibinfo {pages} {1298} (\bibinfo {year} {2020})}\BibitemShut {NoStop}%
\bibitem [{\citenamefont {Head}\ \emph {et~al.}(2003)\citenamefont {Head}, \citenamefont {Levine},\ and\ \citenamefont {MacKintosh}}]{head_deformation_2003}%
  \BibitemOpen
  \bibfield  {author} {\bibinfo {author} {\bibfnamefont {D.~A.}\ \bibnamefont {Head}}, \bibinfo {author} {\bibfnamefont {A.~J.}\ \bibnamefont {Levine}},\ and\ \bibinfo {author} {\bibfnamefont {F.~C.}\ \bibnamefont {MacKintosh}},\ }\bibfield  {title} {\bibinfo {title} {Deformation of {{Cross-Linked Semiflexible Polymer Networks}}},\ }\href {https://doi.org/10.1103/physrevlett.91.108102} {\bibfield  {journal} {\bibinfo  {journal} {Physical Review Letters}\ }\textbf {\bibinfo {volume} {91}},\ \bibinfo {pages} {108102} (\bibinfo {year} {2003})}\BibitemShut {NoStop}%
\bibitem [{\citenamefont {Wilhelm}\ and\ \citenamefont {Frey}(2003)}]{wilhelm_elasticity_2003}%
  \BibitemOpen
  \bibfield  {author} {\bibinfo {author} {\bibfnamefont {J.}~\bibnamefont {Wilhelm}}\ and\ \bibinfo {author} {\bibfnamefont {E.}~\bibnamefont {Frey}},\ }\bibfield  {title} {\bibinfo {title} {Elasticity of {{Stiff Polymer Networks}}},\ }\href {https://doi.org/10.1103/physrevlett.91.108103} {\bibfield  {journal} {\bibinfo  {journal} {Physical Review Letters}\ }\textbf {\bibinfo {volume} {91}},\ \bibinfo {pages} {108103} (\bibinfo {year} {2003})}\BibitemShut {NoStop}%
\bibitem [{\citenamefont {Heussinger}\ and\ \citenamefont {Frey}(2006)}]{heussinger_floppy_2006}%
  \BibitemOpen
  \bibfield  {author} {\bibinfo {author} {\bibfnamefont {C.}~\bibnamefont {Heussinger}}\ and\ \bibinfo {author} {\bibfnamefont {E.}~\bibnamefont {Frey}},\ }\bibfield  {title} {\bibinfo {title} {Floppy {{Modes}} and {{Nonaffine Deformations}} in {{Random Fiber Networks}}},\ }\href {https://doi.org/10.1103/physrevlett.97.105501} {\bibfield  {journal} {\bibinfo  {journal} {Physical Review Letters}\ }\textbf {\bibinfo {volume} {97}},\ \bibinfo {pages} {105501} (\bibinfo {year} {2006})}\BibitemShut {NoStop}%
\bibitem [{\citenamefont {Broedersz}\ \emph {et~al.}(2012)\citenamefont {Broedersz}, \citenamefont {Sheinman},\ and\ \citenamefont {MacKintosh}}]{broedersz_filament-length-controlled_2012}%
  \BibitemOpen
  \bibfield  {author} {\bibinfo {author} {\bibfnamefont {C.~P.}\ \bibnamefont {Broedersz}}, \bibinfo {author} {\bibfnamefont {M.}~\bibnamefont {Sheinman}},\ and\ \bibinfo {author} {\bibfnamefont {F.~C.}\ \bibnamefont {MacKintosh}},\ }\bibfield  {title} {\bibinfo {title} {Filament-length-controlled elasticity in {{3D}} fiber networks},\ }\href {https://doi.org/10.1103/physrevlett.108.078102} {\bibfield  {journal} {\bibinfo  {journal} {Physical Review Letters}\ }\textbf {\bibinfo {volume} {108}},\ \bibinfo {pages} {078102} (\bibinfo {year} {2012})}\BibitemShut {NoStop}%
\bibitem [{\citenamefont {Chen}\ \emph {et~al.}(2023)\citenamefont {Chen}, \citenamefont {Markovich},\ and\ \citenamefont {MacKintosh}}]{chen_nonaffine_2023}%
  \BibitemOpen
  \bibfield  {author} {\bibinfo {author} {\bibfnamefont {S.}~\bibnamefont {Chen}}, \bibinfo {author} {\bibfnamefont {T.}~\bibnamefont {Markovich}},\ and\ \bibinfo {author} {\bibfnamefont {F.~C.}\ \bibnamefont {MacKintosh}},\ }\bibfield  {title} {\bibinfo {title} {Nonaffine {{Deformation}} of {{Semiflexible Polymer}} and {{Fiber Networks}}},\ }\href {https://doi.org/10.1103/PhysRevLett.130.088101} {\bibfield  {journal} {\bibinfo  {journal} {Physical Review Letters}\ }\textbf {\bibinfo {volume} {130}},\ \bibinfo {pages} {088101} (\bibinfo {year} {2023})}\BibitemShut {NoStop}%
\bibitem [{\citenamefont {Ball}(2012)}]{ball_mathematics_2012}%
  \BibitemOpen
  \bibfield  {author} {\bibinfo {author} {\bibfnamefont {J.}~\bibnamefont {Ball}},\ }\href@noop {} {\emph {\bibinfo {title} {Mathematics of {{Liquid Crystals}}}}},\ \bibinfo {type} {Cambridge {{CCA Course}}}\ (\bibinfo {year} {2012})\BibitemShut {NoStop}%
\end{thebibliography}%

\end{document}